\documentclass[prb,aps,showpacs,twocolumn,amsmath,amssymb,floatfix,superscriptaddress]{revtex4-2}

\usepackage[unicode=true, colorlinks=true, citecolor={blue!80!black}, urlcolor={blue!50!black}, linkcolor = {blue!80!black}]{hyperref}
\usepackage{graphicx}
\usepackage{bm}
\usepackage{dsfont}
\usepackage{xcolor}
\usepackage[utf8]{inputenc}
\usepackage{amssymb}
\usepackage{amsmath}
\usepackage[normalem]{ulem}
\usepackage{soul}
\usepackage{physics}

\usepackage{natbib}
\usepackage{hyperref}
\usepackage{float}
\usepackage{sidecap,tikz}
\definecolor{lime}{HTML}{A6CE39}
\DeclareRobustCommand{\orcidicon}{\hspace{-1mm}
	\begin{tikzpicture}
		\draw[lime, fill=lime] (0,0) 
		circle [radius=0.16] 
		node[white] {{\fontfamily{qag}\selectfont \tiny \,ID}};
		\draw[white, fill=white] (-0.0525,0.095) 
		circle [radius=0.007];
	\end{tikzpicture}
	\hspace{-3mm}
}
\foreach \x in {A, ..., Z}{\expandafter\xdef\csname orcid\x\endcsname{\noexpand\href{https://orcid.org/\csname orcidauthor\x\endcsname}
		{\noexpand\orcidicon}}
}

\newcommand{\ftmc}{Departamento de Física Teórica de la Materia Condensada, Instituto Nicolás Cabrera and Condensed Matter Physics Center (IFIMAC), Universidad Aut\'onoma de Madrid, 28049 Madrid, Spain}

\newcommand{\quarc}{Quantum Advanced Research Center (QuARC), Consejo Superior de Investigaciones Científicas (CSIC), Sor Juana Inés de la Cruz 3, 28049 Madrid, Spain}
\newcommand{\icmm}{Instituto de Ciencia de Materiales de Madrid (ICMM), Consejo Superior de Investigaciones Científicas (CSIC), Sor Juana Inés de la Cruz 3, 28049 Madrid, Spain}

\begin{document}
\title{Inductively-protected Andreev (IPA) spin qubit}
\author{J. L. del Olmo N.}
\affiliation{\quarc}\affiliation{\icmm}
\author{F. J. Matute-Cañadas}
\affiliation{\ftmc}
\author{A. Levy Yeyati\orcidC{}}
\affiliation{\ftmc}
\author{R. Seoane Souto\orcidB{}}
\affiliation{\quarc}\affiliation{\icmm}
\author{R. Aguado\orcidA{}}
\affiliation{\quarc}\affiliation{\icmm}
\begin{abstract}
The spin of a quasiparticle trapped in a quantum dot Josephson junction forms the basis of an Andreev spin qubit (ASQ): a semiconductor-superconductor device where the interplay between a localized spin degree of freedom and superconductivity leads to a spin-resolved Josephson potential. In this work, we show that shunting an ASQ with a linear inductor enhances its relaxation time  by separating the spin-qubit states into distinct potential wells in phase space, nearly eliminating wavefunction overlap. The resulting inductively protected Andreev (IPA) spin qubit is equivalent to two fluxoniums in the heavy regime, one for each spin. Thus, the IPA qubit combines the long coherence times, low-frequency ground-state manifold, and large anharmonicity of a protected superconducting qubit with the operational advantages of a spin degree of freedom. 

\end{abstract}
\maketitle
\section{Introduction}
A major obstacle to building a scalable quantum computer \cite{mohseni2025buildquantumsupercomputerscaling} is environmental noise, which causes qubit decoherence through relaxation and dephasing.
Scaling, therefore, demands strategies that mitigate the effects of such noise without sacrificing qubit performance. Beyond quantum error correction \cite{Google-error}, a complementary approach is to engineer the Hamiltonian to inherently protect computational states from local noise~\cite{Danon_protected}.
\begin{figure}
\includegraphics[width=1\linewidth]
{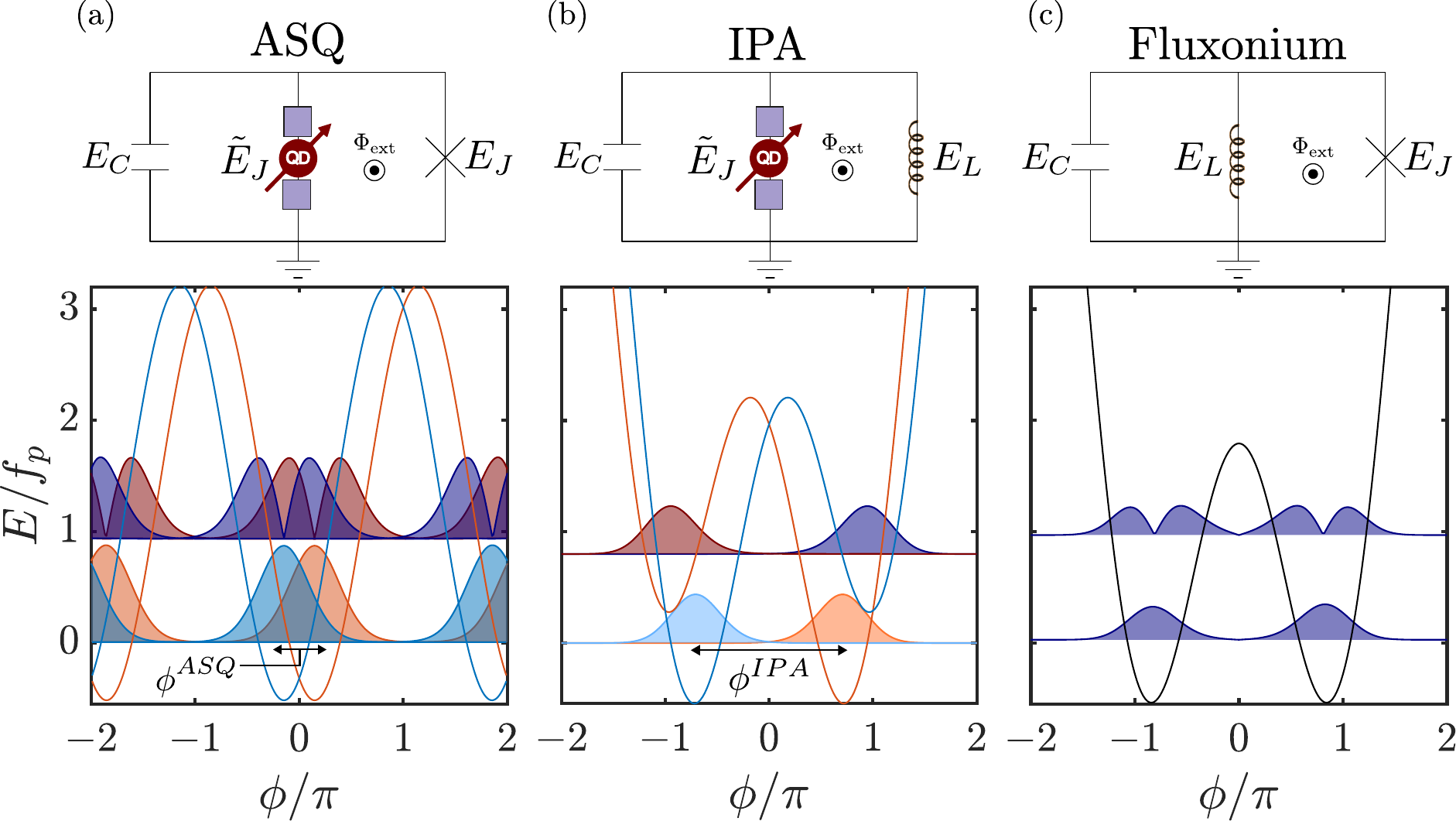}
\caption{Top panels: circuits of the (a) Andreev spin qubit (ASQ) in a transmon setup, (b) inductively-protected Andreev (IPA) spin qubit and (c) fluxonium qubit. Here, $E_C$ is the charging energy, $E_L$ the inductive energy, and $E_J$ and $\tilde{E}_J$ the Josephson energies. 
Bottom panels: Josephson potentials (lines) and wavefunctions (filled curves) of the lowest-energy levels corresponding to the circuits on the top. Blue/red colors correspond to two spin branches. In the ASQ (bottom left), the spin branches
are phase-shifted by a distance $\phi_0^{\rm ASQ}$, see Eqs.~(\ref{QD-JJ}-\ref{spin-branches}). Due to the periodic boundary conditions, the phase separation between spin wells is $\phi^{\rm ASQ} \leq \pi$. In the IPA case (bottom center),
the inductor lifts the periodicity and raises the external wells in energy, allowing for a separation between spins up to $\phi^{\rm IPA}\lesssim 2\pi$. The IPA potential at low energies is reminiscent to that of a fluxonium biased at half-flux $\Phi_{\rm ext}=\Phi_{0}/2$ (bottom right), but featuring uncoupled spin-resolved wave functions localized at different minima in phase space.}
\label{fig1}
\end{figure}

Apart from topological qubits, which encode the Hilbert space in global degrees of freedom to ensure protection against local fluctuations~\cite{Kitaev2001,kitaev2003fault,nayak2008,Aguado-KouwenhovenPT2020,Pino_PRB2024,Pan_PRB2025}, there are other ways to design protected qubits with superconducting circuits. By engineering multiple degrees of freedom and modifying the periodicity of Josephson potentials it is possible to improve the trade-off between bit-flip and dephasing suppression \cite{Gyenis2021,Danon2021,Siddiqi2021}.  Examples of this second approach include the $\cos 2{\phi}$ parity protected qubits~\cite{cos2phi2026}, the bifluxon qubit \cite{PRXQuantum.1.010307},  the $0-\pi$ qubit \cite{PhysRevA.87.052306}, the harmonium qubit \cite{dd96-gcb6} or the GKP qubit \cite{PhysRevX.15.011011}.

Here, we explore these ideas in the context of the recently developed Andreev spin qubit (ASQ) \cite{Hays2021,PitaVidal2023,nazarov-manipulation}.
The ASQ consists of two opposite spin states embedded within a superconducting circuit, which affords a compact footprint and leverages the advantages of circuit quantum electrodynamics, enabled by the coupling between spin and supercurrent  (e.g., Fig.~\ref{fig1}a). However, its practical utility is presently limited by short coherence times. We demonstrate that shunting the ASQ with a capacitor and an inductor in a fluxonium-like architecture \cite{Manucharyan2009} effectively mitigates some of its limitations by enabling higher lifetimes of trapped spin. The inductive element increases the disjointness of the qubit wavefunctions in phase space. This defines a new, robust design which we term inductively protected Andreev (IPA) qubit. Conceptually, the IPA is a low-frequency qubit similar to a heavy fluxonium \cite{PhysRevX.9.041041,PhysRevX.11.011010} biased 
close to half-flux $\Phi_{\rm ext}=\Phi_{0}/2=\frac{h}{4e}$, 
with phase-separated wavefunctions, nearly degenerate states, and a large anharmonicity. In contrast to the fluxonium, the IPA features opposite spins in each of the minima of the double well phase potential, effectively suppressing direct spin transitions, see Fig.~\ref{fig1}(b,c).
Our analysis reveals an enhancement in the relaxation time  $T_1$ and overall coherence, as demonstrated by benchmarks against the standard fluxonium and ASQ. While recent works have begun to explore circuit improvements for ASQs, like e.g circuits with superconducting islands (transmon-like) \cite{kurilovich2025andreevspinqubitprotected,Tjernshaugen2026coherentcontrolspinmons} and multiterminal devices \cite{manesco2026looplessmultiterminalquantumcircuits}, the IPA design provides further protection against relaxation. 
This, together with its operational benefits, such as enhanced anharmonicity and magnetic field tunability, suggests that the IPA is an interesting geometry worth exploring in  existing semiconductor-based fluxonium implementations \cite{PhysRevApplied.14.064038,PRXQuantum.6.010326,isakov2026}.

The remainder of this paper is organized as follows. Section~\ref{concept} introduces the concept of spin protection provided by the inductor. Section~\ref{Sec.QubitProperties} details the qubit properties, covering the harmonic approximation (Sec.~\ref{harmonic}), wavefunction separation in phase space (Sec.~\ref{subsec:Wavefunction}), wavefunction overlaps (Sec.~\ref{subsec:overlaps}), anharmonicity (Sec.~\ref{subsec:anharmonicity}), flux and magnetic field dependence (Sec.~\ref{subsec:flux} and Sec.~\ref{magnetic field}), and qubit operations (Sec.~\ref{subsec:qubitOperations}). In Sec.~\ref{decoherence-section}, we analyze the characteristic $T_1$ and $T_2$ times of the IPA  and benchmark them against the ASQ and the fluxonium. Finally, we present the conclusions and outlook in Sec.~\ref{Sec:conclusions}.

\section{Concept \label{concept}}
Our starting point is a hybrid 
Josephson junction (JJ) with superconducting phase difference $\phi$ based on a semiconducting quantum dot (QD) in a doublet ground state (odd-fermionic parity) \cite{pitavidal2025novelqubitshybridsemiconductorsuperconductor}. When considering spin-orbit (SO) coupling, such QD JJs can be described by the following 
Hamiltonian~\cite{Padurariu2010,Bargerbos2023b}
\begin{equation}
\label{QD-JJ}
    H^{QD}_J({\phi}) = E_0 \cos{({\phi})} \sigma_0 - E_{\rm SO} \sin{({\phi})}
    \vec{\sigma} \cdot\vec{n}_{\rm SO}\,.
\end{equation}
Equation~\eqref{QD-JJ} differs notably from the standard Josephson potential term that enters in transmon qubit models $H_J({\phi})=-E_J\cos{({\phi})}$, leading to novel physics and functionalities that we describe in what follows.
The first term, proportional to $\cos{\phi}$ is minimized at phase $\phi=\pi$($E_0>0$)  due to the doublet ground state ($\pi$-junction regime)~\cite{Bargerbos2022}. This is in contrast with that of the conventional tunnel junction, which with an even parity ground state, is minimized at $\phi=0$ \footnote{Although $E_0>0$ describes a $\pi$-junction regime, the model admits other situations, including $E_0\approx 0$ and $E_0<0$, depending on microscopic details of the QD junction such as the amount of spin-flip tunneling and the contribution from tunneling through higher levels. In fact, fittings to two-tone spectroscopy data confirm that indeed such a variety of situations exists in the experiments \cite{Bargerbos2023b}}. The second term in Eq.~\eqref{QD-JJ}, proportional to $\sin{\phi}$, is a spin-dependent contribution to the Josephson coupling originated from the SO coupling in the direction $\vec{n}_{\rm SO}$, with Pauli matrices $\vec{\sigma} = (\sigma_x,\sigma_y,\sigma_z)$ acting in spin space \footnote{The model in Eq.\eqref{QD-JJ} can be obtained analytically from a so-called superconducting Anderson impurity model in the limit of large superconducting gap and including extra terms  taking into account cotunneling through higher levels as well as spin-flip tunneling. For full details see the supplemental information in Ref. \cite{Bargerbos2023b}.}. The SO in the QD JJ directly induces a phase-dependent lifting of spin degeneracy, even in the absence of a Zeeman field, thereby linking spin and superconducting phase degrees of freedom. Combined with the voltage-tunable Josephson energies, $E_0$ and $E_{\rm SO}$, this linking enables circuit QED implementations with novel gate and flux control capabilities \cite{pitavidal2025novelqubitshybridsemiconductorsuperconductor,Bargerbos2023b,PitaVidal2023,PitaVidal2024}.

The eigenvalues of Eq.~\eqref{QD-JJ} define spin-dependent Josephson potentials along the SO direction (in what follows, we fix the SO axis along the $x$ direction) given by
\begin{equation}
\label{spin-branches}
U_{\pm}^{QD}(\phi)=E_0 \cos(\phi)\pm E_{SO}\sin(\phi)=-\tilde{E}_J\cos(\phi-{\phi}_{\pm})\,,
\end{equation}
with amplitude $\tilde{E}_J=\sqrt{E_0^2+E_{SO}^2}$ and minima located at 
\begin{equation}
 \label{phase-minima}{\phi}_{\pm}= \pi\pm\phi_0=\pi  \pm \arctan(E_{SO}/E_0)\,,
\end{equation}
for the up and down spin, module $2\pi$. Such displaced minima allow to define an effective separation between spin wells $\phi^{\rm ASQ}=\min \{|\phi_{+}{-}\phi_{-}|, 2\pi-|\phi_{+}{-}\phi_{-}| \} \leq \pi$.

Conversely, when the junction is inductively shunted, the potential acquires a parabolic term,
$U_L({\phi}) = \frac{1}{2}E_L \phi^2$ which, including the external flux, results in a total potential
\begin{equation}
\label{Josephson-IPA}
U^{\rm{IPA}}_{\pm}(\phi)=U_{\pm}^{QD} (\phi - \phi_{\rm{ext}}) + U_L(\phi)\,,  
\end{equation} that is no longer $2\pi$-periodic. Due to this inductive term, the potential wells outside the $[-\pi,\pi]$ range raise in energy, allowing for localized wavefunctions with separation $\phi^{\rm IPA} \sim2\pi > \max \{\phi^{\rm ASQ}\}$, see Fig~\ref{fig1}b (bottom). The resulting Josephson double-well potential is reminiscent of those of other protected qubits, such as the $\cos2{\phi}$ or the fluxonium biased at half-flux, see Fig~\ref{fig1}(c) (bottom). However, in the case of the IPA, the two spin sectors are completely uncoupled in the absence of a magnetic field, featuring spin-resolved separation in phase space. The consequences of this Josephson potential in a superconducting qubit circuit are discussed below.

\section{Qubit Properties}
\label{Sec.QubitProperties}

For benchmarking the IPA qubit properties, it is convenient to compare it to other qubits like the ASQ and the fluxonium. The case of an ASQ in a SQUID transmon setup, top of panel in Fig.~\ref{fig1}(a), as implemented in Refs.~\cite{Bargerbos2023b,PitaVidal2023}, can be modeled by the Hamiltonian
\cite{Bargerbos2023b}
\begin{equation}
\label{asq}
    H^{ASQ}=\left[4E_Cn^2 -E_J \cos\phi\right] \sigma_0 +H^{QD}_J(\phi-\phi_{\rm ext}),
\end{equation}
where $n$ is the Cooper pair number operator, conjugated to the phase operator $\phi$, $E_C$ is the charging energy, which is mainly determined by the superconducting island, $E_J$ is the Josephson coupling of the tunnel junction, and
$\phi_{\rm ext}=2\pi\Phi_{\rm ext}/
\Phi_0$ is the external flux bias measured in units of the superconducting flux quantum $\Phi_0=h/2e$. 

Following the discussion in Sec.~\ref{concept}, we modify the above ASQ model by shunting the QD JJ by an inductor, see~\ref{fig1}(b). The IPA qubit model thus reads

\begin{equation}
    H^{\rm IPA}=\left[4E_C n^2 + \frac{1}{2}E_L\phi^2\right] \sigma_0 +H^{QD}_J(\phi-\phi_{\rm ext}),
    \label{H-IPA}
\end{equation}

\noindent
where the charging energy in this case would be dominated by the inductor capacitance in a nanowire junction implementation \cite{caceres2026}.
Finally, a conventional fluxonium, Fig.~\ref{fig1}(c) top, is described by the spinless Hamiltonian

\begin{equation}
\label{flux-qubit}
H^{\rm flux}=4E_Cn^2 + \frac{1}{2}E_L\phi^2 - E_J\cos({\phi}-\phi_{\rm ext}).
\end{equation}

\noindent
In conventional tunnel junctions, the charging energy of the circuit is dominated by the junction intrinsic capacitance, and $\phi_{\rm ext}$ drop takes place on the inductive term. The intrinsic capacitance of weak links, though, is substantially reduced and we expect most of the phase drop there~\cite{caceres2026}. Notice, however, that the particular phase drop distribution only affects time dependent observables, but not the energy spectrum \cite{You2019,Bryon2023}.

The fluxonium qubit features two characteristic regimes with different protection against noise. In the so-called ``light regime'', with  $E_C > E_J\gg E_L$, the energy levels have little dispersion against $\phi_{\rm ext}$, suppressing flux-noise dephasing as the qubit spectrum becomes harmonic, approaching a plasma frequency  $f_p^L=\sqrt{8E_CE_L}$ (hereafter, we use $h=1$). The opposite, so-called ``heavy regime'', with  $E_J \gg E_L \sim E_C$ minimizes the probability of bit-flip events, hence enhancing relaxation times. We here exploit a similar idea to the fluxonium in the heavy regime but with the extra twist of having spin-resolved separation in phase space for $E_{SO}\neq 0$, see Fig.~\ref{fig1}(b). Importantly, this idea exploits that fact that the IPA model in Eq.~\eqref{H-IPA} is \emph{equivalent to a fluxonium for each spin}, with $E_J \rightarrow \tilde{E}_J$ and $\phi_{\rm ext} \rightarrow \phi_{\rm ext} + \phi_{\pm}$. Therefore, reducing the states overlap in phase space provides additional protection to the spin qubit with respect to various noise sources.  

In the benchmarking procedure that follows, we note that, when  comparing the IPA qubit against the ASQ, we set $\phi^0_{\rm ext}=0$ for both. In contrast, we fix the reference flux for the fluxonium qubit 
at $\phi^0_{\rm ext}=\pi$. We define flux detunings as 
$\delta \phi_{\rm ext}=\phi_{\rm ext} - \phi_{\rm ext}^0$. Thus $\delta \phi_{\rm ext}=0$ is associated with a double-well potential configuration in all three cases, see Fig.~\ref{fig1}.

\subsection{\label{harmonic}Qubit spectrum in the harmonic limit}
\begin{figure}[!htbp]
    \centering
\includegraphics[width=0.45\textwidth]{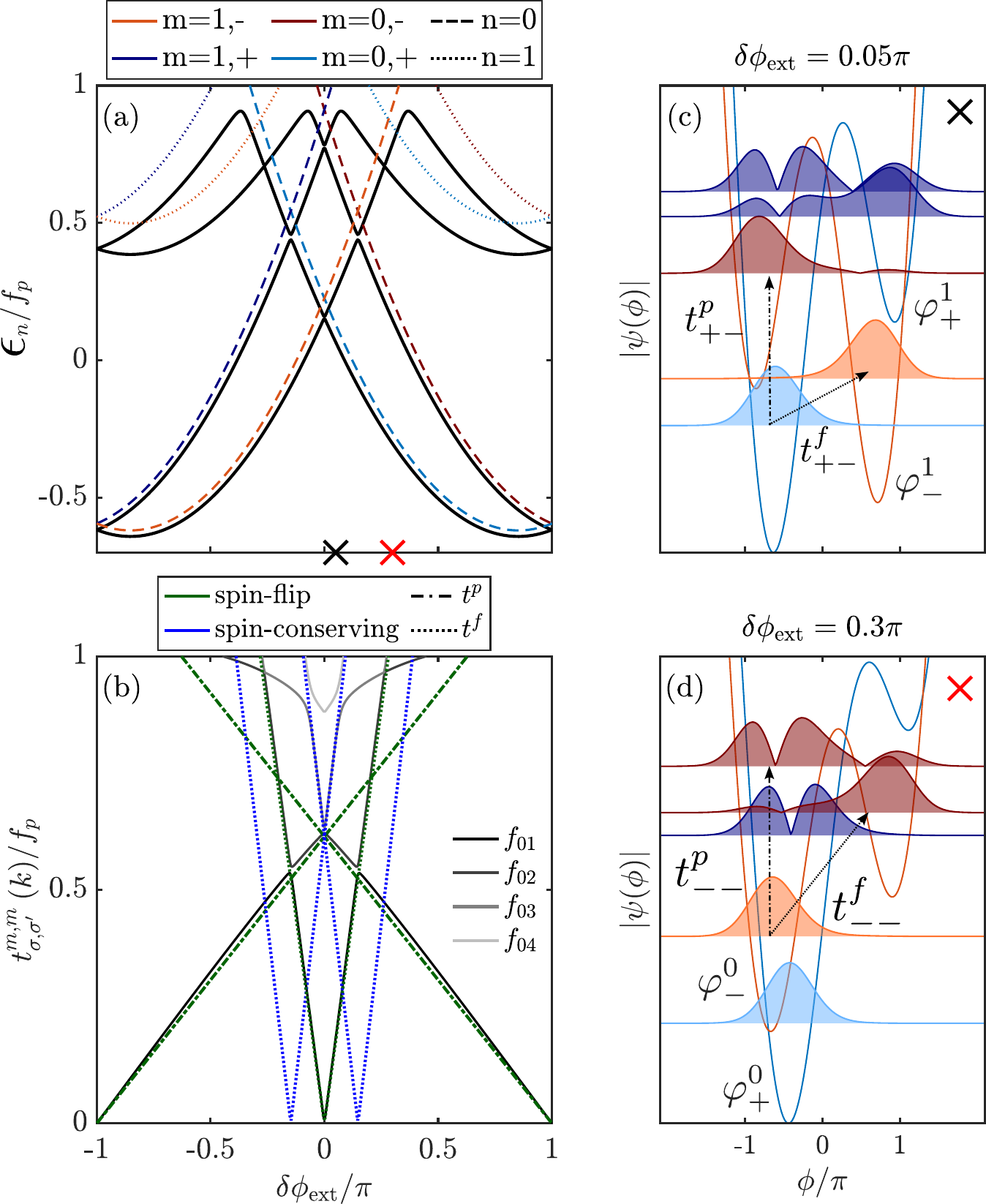}
\caption{(a) Comparison between the eigenvalues of the IPA qubit, obtained from  the exact diagonalization of Eq.~\eqref{H-IPA} (black lines), and the harmonic approximation, Eq.~\eqref{approx-harmonic-energy} (dashed and dotted lines). (b) Corresponding transition energies,  see Eq.~\eqref{transitions-harmonic} for the harmonic approximation in dashed. (c,d) Wave functions and spin-resolved Josephson potential normalized by plasma frequency $f_p$, for $\delta\phi_{\rm{ext}}=0.05\pi$ (c) and $\delta\phi_{\rm{ext}}=0.3\pi$ (d). The arrows illustrate plasma and fluxon-like spin-flip with $k=0$ in (c) and spin conserving transitions in (d) c.f. Eq. \eqref{transitions-harmonic}. In this latter case, the plasma one presents $k=1$ and the fluxon $k=0$. Parameters: $\tilde{E}_J/E_C=11.11$ and $E_L/E_C=2.78$, corresponding to $\chi\approx 1.25$.
}
\label{fig2}
\end{figure}

Before discussing the full solution to the qubit model in Eq. \eqref{H-IPA}, it is useful to build some physical intuition by analyzing the limit where the Josephson energy dominates over the charging (kinetic) energy, such that the wavefunctions are well-localized in phase space. In this limit, the minima $\phi^m_\pm$ of $U^{\rm IPA}_\pm(\phi)$ can be approximated from the minima $\varphi^m_\pm$ of the sinusoidal term of the potential,
\begin{equation}
\label{Eq:phi_pm_harmonic}
\begin{gathered}
    \varphi_{\pm}^m= 2\pi \left( m \pm \frac{1}{2}-\frac{1}{2} \right) +  \delta\phi_{\rm ext} \ \mp(\pi -\phi_0),
\end{gathered}
\end{equation}
where $m$ labels the fluxon well and $\phi^m_\pm  \approx  \varphi^m_\pm/\chi$, with $\chi=1+\frac{E_L}{\tilde{E}_J}$. Near each minima, the potential in Eq.~\eqref{Josephson-IPA} can be Taylor expanded as 
\begin{equation}
\label{harmonic-IPA}
 U^{\rm{IPA}}_{\pm}(\phi)\approx U^m_{\pm} + \frac{1}{2}E^\chi_J\left(\phi-\phi_{\pm}^m\right)^2, 
\end{equation}
an approximation that is valid for small values of $E_L/\tilde{E}_J$ and $|m|$. Here, the curvature is given by $E^\chi_J=\tilde E_J\chi$. The height of potential at those minima is

\begin{equation}
\label{harmonic-well}
    U^m_{\pm} \approx -\tilde{E}_J + \frac{E_L}{2\chi} \left(\varphi_{\pm}^m\right)^2,
\end{equation}
\noindent 

\noindent

The harmonic approximation in each well gives  \textit{plasma} excitations $n=0,1,2...$ with energies 
\begin{align}
\label{approx-harmonic-energy}
    \varepsilon^{m,n}_{\sigma} \approx U^m_{\sigma} + f^\chi_p\left(n+\frac{1}{2}\right),
\end{align}
where, owing to the inductor, the bare plasma frequency $ f_p = \sqrt{8 E_C \tilde E_J}$ gets renormalized 
and becomes $f^\chi_p = \sqrt{8 E_C E^\chi_J} =\sqrt{\chi}f_p$. Consequently, the zero-point fluctuations of the harmonic potential in Eq. \eqref{harmonic-IPA} read $\phi_{\rm zpf}^\chi = (2E_C/E_J^\chi)^{1/4}$. 

Figure \ref{fig2}(a) presents the low-energy spectra of the IPA in the heavy regime as a function of the external flux detuning, given by Eq.~\eqref{H-IPA}. We may identify the spin (red and blue colors), the fluxon number (dark and light tonalities), and the plasmon modes (dashed and dotted) within the harmonic approximation in Eq.~\eqref{approx-harmonic-energy}.

Figures \ref{fig2}(c) and \ref{fig2}(d) show the associated potential energy and wavefunctions for two values of $\delta\phi_{\rm ext}$. Since the system can be understood as two spin-dependent copies of a fluxonium (orange and cyan), the lowest minimum of each spin branch around the symmetry point $\delta\phi_{\rm{ext}}=0$ corresponds, respectively, to the fluxon state $m=0$ (chosen to be the left side of the double-well potential) or $m=1$ (right side), see labels near the wells. Assuming uncoupled wells, the lowest wavefunctions located around these minima, read

\begin{equation}
\label{eq:general states wf}
    \psi_{\sigma}^{m,n=0}(\phi) =  A \exp \left[ - \kappa \left( \phi - \phi_{\sigma}^m \right)^2 \right],
\end{equation}
with $A= \left (\frac{2\kappa} {\pi}\right)^{1/4}$ and $\kappa=\frac{\sqrt{\chi}}{4\phi_{\rm zpf}^2}$. Specifically, $\psi_{+}^{0,0}(\phi)$ and $\psi_{-}^{1,0}(\phi)$ correspond to the two lowest-energy states that form the qubit basis.

Using Eq.~\eqref{approx-harmonic-energy}, we can approximate expressions for the transitions 
$t^{m,m'}_{\sigma,\sigma'} \left(k=\left|n-n'\right|\right)=|\varepsilon^{m}_{\sigma}(n)- \varepsilon^{m'}_{\sigma'}(n')|$ as:

\begin{equation} \label{transitions-harmonic}
t^{m,m'}_{\sigma,\sigma'}(k) = \frac{E_L}{2\chi}\left[ \left(\varphi_{\sigma}^m\right)^2 - \left(\varphi^{m'}_{\sigma'}\right)^2 \right] + f^\chi_pk .
\end{equation}

Within each spin fluxonium, we can classify the corresponding spin-conserving transitions of lower energy into those that shift the fluxon well, $t^f_{\sigma ,\sigma}(k) \equiv t^{m, m+1}_{\sigma,\sigma}(k)$, dotted blue lines in Fig.~\ref{fig2}(b), and those that excite a higher energy state of the same fluxon, $t^p_{\sigma ,\sigma}(k)\equiv t^{m,m}_{\sigma,\sigma}(k)$, see labels in Fig.~\ref{fig2}(d). There are also transitions that involve spin-flip, in wells that are either nearby or separated in phase by $\lesssim 2\pi$ (depending on $\phi_0$). These transitions are crucial for qubit operations, see Sec.~\ref{subsec:qubitOperations} for a discussion. Spin-flip transitions between neighboring wells with different spins are zero at $\delta\phi_{\rm ext}=\pm\pi$, being approximately $\mp2E_L\phi_0(\pi\pm\delta\phi_{\rm ext})/\chi$ with $\delta\phi_{\rm ext}\in[-\pi,\pi]$, see dashed-dotted lines in Fig.~\ref{fig2}(b). On the other hand, spin-flip transitions between different wells are zero at $\delta\phi_{\rm ext}=0$, with
\begin{eqnarray}
\label{qubit-freq-harmonic}
f^{\rm IPA}_{01}\approx t^f_{+,-}(0)&=&\frac{E_L}{2\chi}\left[ \left(\varphi_{+}^{0}\right)^2 - \left(\varphi_{-}^1\right)^2 \right]\nonumber\\
&=&2\frac{E_L}{\chi}\left|\delta\phi_{\rm{ext}}\right|\left|\pi-\phi_{0}\right|,
\end{eqnarray}
for $E_0,E_{SO}>0$, $\delta\phi_{\rm ext}\in[-\pi,\pi]$. Equation~\eqref{qubit-freq-harmonic} determines the IPA qubit frequency close to $\delta\phi_{\rm ext}=0$, Fig.~\ref{fig2}(c). 

\subsection{Wavefunctions in phase space}
\label{subsec:Wavefunction}
Moving beyond the previous approximation in the heavy regime, we now provide a broader analysis of the lower energy wavefunction, which form the IPA qubit at $\delta\phi_{\rm ext}=0$. The system has three independent parameters in Eq.~\eqref{H-IPA}: the separation of the spin branches ($\phi_{\pm}$), the height of the wells ($\tilde{E}_J/E_C$), and the aperture of the inductive parabola ($E_L/E_C$).
\begin{figure*}
    \centering
    \includegraphics[width=0.85\textwidth]{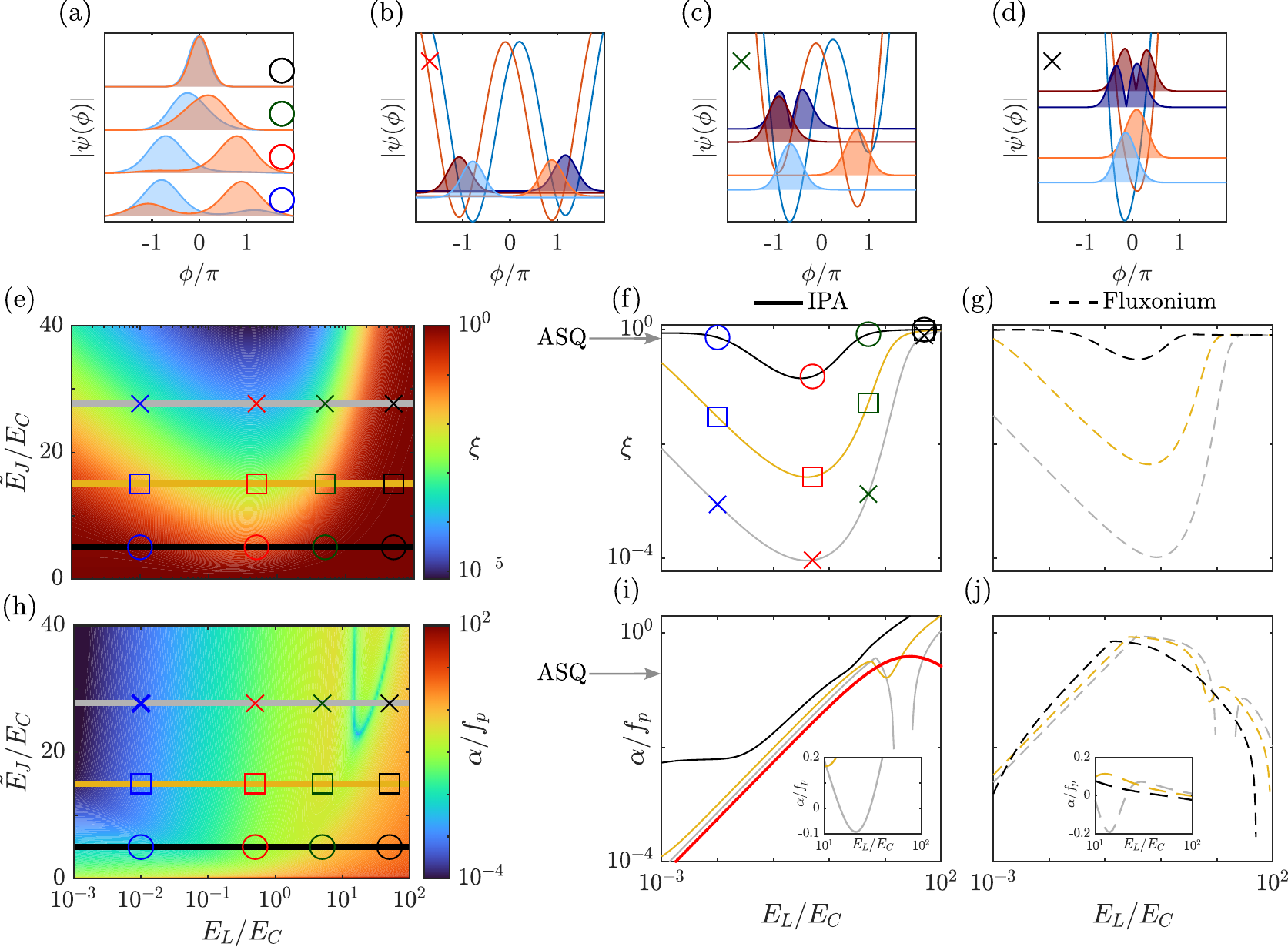}
    
\caption{(a) Lowest-energy eigenstates of the IPA qubit, Eq.~\eqref{H-IPA}, for $\tilde{E}_J/E_C=5$ and $E_L/E_C = 50,\; 5,\; 0.5$, and $0.01$ from top to bottom, where red/blue denote spin up/down along the spin-orbit direction. (b)--(d) Josephson potential (solid lines) and lowest-energy wavefunctions, $|\psi_j(\phi)|$, (colored curves) for $\tilde{E}_J/E_C\gg1$, and $E_L/E_C = 0.5,\; 5,\;  50$. (e) Wavefunction overlap ($\xi$), as defined in Eq.~\eqref{eq:overlap_def}, as a function of $E_L/E_C$ and $\tilde{E}_J/E_C$. Grey, yellow, and black lines are cuts for $\tilde{E}_J/E_C=27.8$, $15$, and $5$ while the symbols represent situations with $E_L/E_C = 50,\; 5,\; 0.5,$ and $10^{-2}$. (f) $\xi$  as a function of $E_L/E_C$ for the $\tilde{E}_J/E_C$ lines shown in panel (e). (g) $\xi$ for fluxonium qubits along the same lines of $\tilde{E}_J/E_C$ as in panel (f). In both cases, IPA (panel f) and fluxonium (panel g) we use $\delta\phi_{\rm ext}=0.05\pi$, that induces phase localization of the qubit wavefunctions, allowing a better comparison between the qubits. We also mark the wavefunction overlap for the ASQ and $\tilde{E}_J/E_C=27.8$, corresponding to the minimum of the three considered cases, as an arrow. (h) Anharmonicity, $\alpha$ in Eq.~\eqref{anharmon}, as a function of $E_L/E_C$ and $\tilde{E}_J/E_C$. (i) shows cuts corresponding to the horizontal coloured lines in panel (h), the red line represents the result from the harmonic approximation $\alpha=2\pi\phi_0E_L/\chi$, for the same parameters as the grey line, and the arrow is the anharmonicity for $\tilde{E}_J/E_C=27.8$. (j) same as (i) for fluxonium. Same fluxes as in panels (f) and (g). The insets in both panels are zooms in linear scale showing regions with negative anharmonicity. The QD parameters are fixed to $E_0/E_{SO}=2$. To interpret our results in terms of physical units, it is useful to recall typical values for the ASQ~\cite{PitaVidal2023}. The charging energy in frequency units is $E_C\sim 300$ MHz, while the Josephson terms, $E_0$ and $E_{SO}$, are typically in the GHz range.}
       \label{fig3}
\end{figure*} 

The interplay between some of these parameters is illustrated in Fig.~\ref{fig3}(a), which shows the effect of the inductive energy on the two lowest-energy wavefunction amplitudes for the $+/-$ spins (cyan/orange shades), for $\tilde{E}_J/E_C=5$ and decreasing $E_L/E_C$ ratio from top to bottom. In the limit $E_L/E_C\gg 1$ (black circle), the inductive term dominates the Josephson potential such that the low-energy physics is that of two almost-overlapping spin-resolved harmonic potentials, see also Fig.~\ref{fig3}(d). In this limit, the spin states are localized near $\phi\approx\delta\phi_{\rm{ext}}=0$, resulting in a negligible phase separation between spin branches. In a heavy fluxonium regime, $E_L/E_C\lesssim 1$ , see {\it e.g.} the case with $E_L/E_C=0.5$, red circle in Fig.~\ref{fig3}(a) and Fig.~\ref{fig3}(b), the two spin wavefunctions separate in phase space by almost $2\pi$. As the fluxonium becomes lighter, for $E_L/E_C <1$, blue circle in Fig.~\ref{fig3}(a), the wavefunctions start to delocalize in phase space beyond their original well. Such a superinductive regime is detrimental to the separation between the IPA states, though in the even-parity Andreev sector it can be used to define another protected regime, dubbed the ``FerBo qubit'' \cite{caceres2026}. Finally, the Blochnium limit  with $E_L/E_C \ll 1$ and $\tilde{E}_J\approx E_C$ \cite{Blochnium2020}, in which the wavefunctions live over several wells of the potential, further deteriorating the disjointness in phase, though reducing the dephasing by $\phi_{\rm ext}$ noise and representing another route to protected qubit design, though it lies beyond the scope of this work.

The previous discussion describes the existence of an optimal regime in $E_L/E_C$ for the spin separation. Such a sweet spot also exists when tuning the relation between $E_0$ and $E_{SO}$, i.e., $\phi_0$ at a fixed $\tilde{E}_J$, which is described in App.~\ref{appendix_harmonic_approx}. The third independent parameter, $\tilde{E}_J/E_C$, monotonously reduces the the wavefunction overlap by making the regime heavier, as will be analyzed below.

\subsection{Overlaps}
\label{subsec:overlaps}
To gain insight about the qubit performance, we now focus on the wavefunction \textit{absolute} overlap, defined as
\begin{equation}
    \xi=\int d\phi\, |\psi_0(\phi)|\,|\psi_1(\phi)|,
    \label{eq:overlap_def}
\end{equation}
with $\psi_{1,0}(\phi)$ being the wavefunctions of the two lowest-energy states of Eq.~\eqref{H-IPA}, which correspond to the $\pm$ spins. This overlap measures disjointness in general but, in particular, it coincides with the matrix element of the magnetic noise perpendicular to the SO, $\partial H /\partial B_\perp$ (see subsection \ref{relaxation-decoherence}). For benchmarking, we compare with the ASQ and fluxonium cases.

The absolute overlap, features distinct regions  differing by various orders of magnitude, as illustrated in  Fig.~\ref{fig3}(e) where $\xi$ is represented as a function of $E_L/E_C$ and $\tilde{E}_J/E_C$. The overlap is found to decrease exponentially with increasing $\tilde{E}_J/E_C$, a direct consequence of wavefunction localization: as $\tilde{E}_J$ grows, the potential barrier separating the two minima increases, confining the spin states and exponentially suppressing their mutual overlap \footnote{The spin-flip transitions can be further suppressed by the 
Franck-Condon principle, since the flip of the quasiparticle spin can only happen if it is accompanied by the excitation of plasmons \cite{kurilovich2025andreevspinqubitprotected}. Although we do not consider such mechanism here, we expect it will also lead to further suppression in the IPA regime.}. As discussed before, in the $E_L/E_C$ axis there is a sweet spot around $E_L \sim E_C$, across which $\xi$ decreases by several orders of magnitude, see the blue region in Fig.~\ref{fig3}(e). This region defines the IPA regime.

The significant reduction in overlaps is better illustrated in the linecuts of Fig.~\ref{fig3}(f), taken at fixed $\tilde{E}_J/E_C$ values along the horizontal lines in Fig.~\ref{fig3}(e). For comparison, we also show results for the fluxonium qubit in Fig.~\ref{fig3}(g). First, absolute overlap reaches a minimum for $E_L/ E_C\sim 1$ for the IPA and the fluxonium. This minimum is similar in both cases, illustrating a similar wavefunction localization. Interestingly, the minimum overlap in the IPA is several orders of magnitude smaller that the one in the ASQ case, illustrating an enhanced wavefunction separation provided by the inductor, see the arrow in Fig.~\ref{fig3}(f).

\begin{figure}
    \centering
\includegraphics[width=0.45\textwidth]{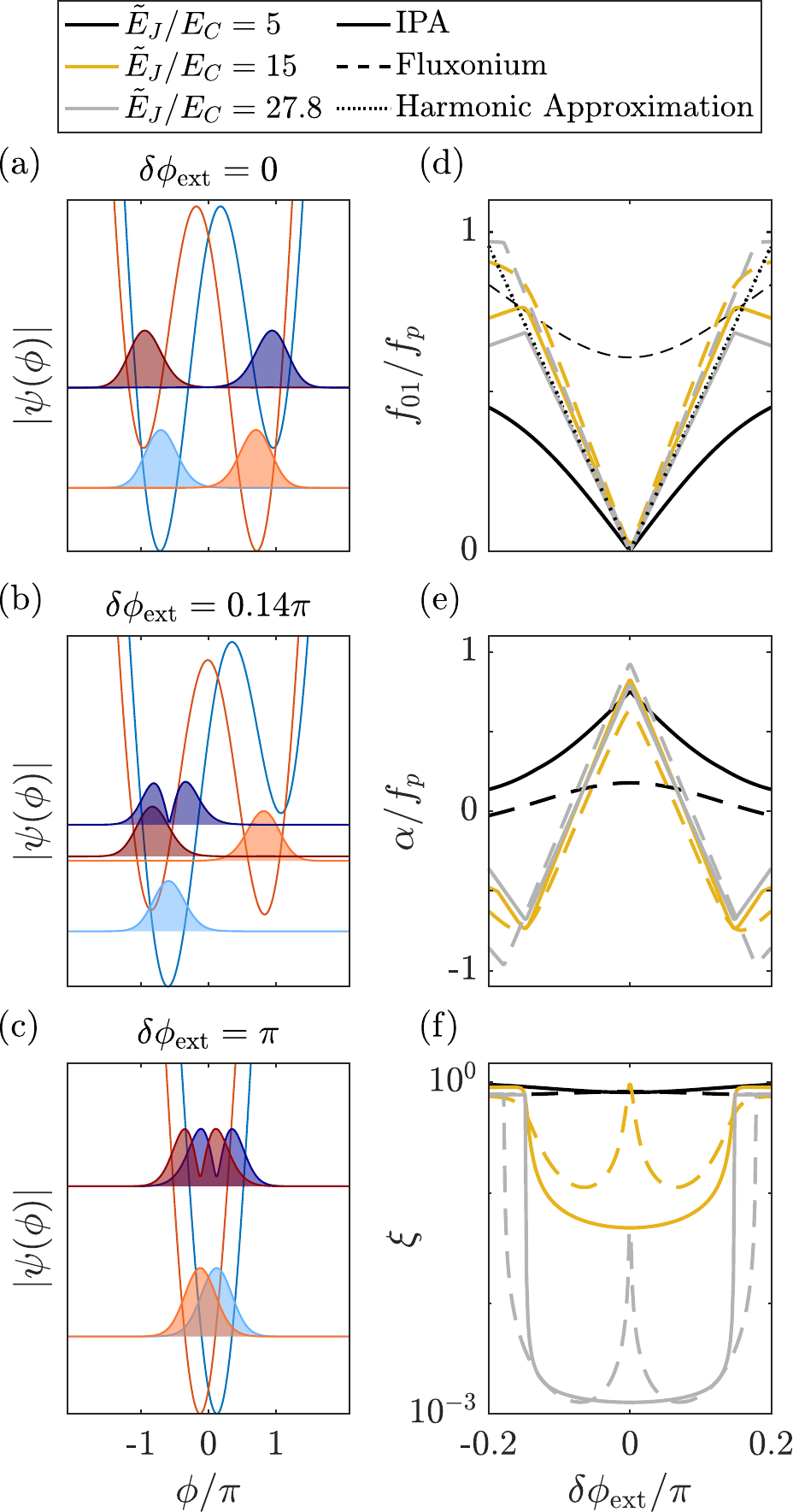}
\caption{Left panels: Josephson potential (thick lines) and lowest-energy eigenstates and wavefunctions (colored curves) for  $E_L=5E_C$, $\tilde{E}_J/E_C=27.8$, $E_{0}=2E_{SO}=2$, and different applied fluxes (a) $\delta\phi_{\rm{ext}}=0$, (b) $\delta\phi_{\rm{ext}}=0.14\pi$ and (c) $\delta\phi_{\rm{ext}}=\pi$.
At $\delta\phi_{\text{ext}}\approx0.14\pi$, a level crossing occurs for the chosen parameters. Below, this value, the two lowest-energy states belong to well-separated wells with opposite spins. Above, the two lowest-energy states no longer belong to distinct wells. While the exact flux value for a level crossing depends on the chosen parameters, the crossing itself reflects a general feature where spin states no longer belong to well-separated wells, compare panels (a) and (c). Right panels: (d),(e) and (f) are the qubit frequency $f_{01}$, anharmonicity $\alpha$, respectively, and wavefunction overlap $\xi$ for $E_L=5E_C$. Solid (dashed) lines correspond to the IPA (fluxonium) for different Josephson couplings $\tilde{E}_J/E_C$ ($E_J/E_C$). The dotted line in panel (d) corresponds to the harmonic approximation given in Eq.~\eqref{qubit-freq-harmonic} for $\tilde{E}_J/E_C=27.8$.}
\label{fig4}
\end{figure}

Here, it is important to emphasize that a critical tradeoff arises from the interplay between wavefunction localization and energy detuning, since the qubit frequency increases as one moves away from $\delta\phi_{\rm ext}=0$, see also Fig.~\ref{fig4}. While deviating from $\delta\phi_{\rm ext}=0$ rapidly amplifies flux-noise-induced dephasing, thereby reducing $T_2$, it simultaneously enhances the relaxation time $T_1$. In the fluxonium, this improvement stems from suppressed dielectric $\sim 1/f_{01}$ and inductive losses $\sim 1/f^3_{01}$, respectively, as well as reduced charge and flux matrix elements  $\langle 0|  n | 1 \rangle =\frac{f_{01}}{8E_C}\langle 0 | \phi | 1\rangle $ \cite{PhysRevX.9.041041,PhysRevX.11.011010}. In the IPA case, this $T_1$ enhancement is even stronger since the main relaxation mechanism due to magnetic noise is supressed by the spin separation in phase space, see Sec.~\ref{decoherence-section}.

Using an extension of the harmonic approximation presented in Sec.~\ref{harmonic} that uses four wells, (see Appendix \ref{appendix_harmonic_approx}), we derive an expression for the overlap, $\xi^h$, that captures the low-energy physics of the qubit near the IPA regime as

\begin{gather}
\label{eq_overlaph_approx_main}
\xi^{h} \sim  e^{-\frac{1}{2\chi^{3/2}}\left(\frac{\pi -\phi_0}{\phi_{\rm zpf}}\right)^2 }+\frac{E_S}{\left|h_\sigma\right|} e^{-\frac{1}{2\chi^{3/2}}\left(\frac{\phi_0}{\phi_{\rm zpf}}\right)^2 }\,,
\end{gather}

\noindent
where $E_S \approx 4 \left( \frac{8\tilde{E}_J^3 E_C}{\pi^2} \right)^{1/4} e^{-\sqrt{8\tilde{E}_J/E_C}}$ is the phase slip amplitude 
and 
$h_{\sigma}= \pi E_L(\delta \phi_{\rm ext}+\sigma\phi_0)/\chi$ is the height difference between the wells with spin $\sigma$. This expression is able to reproduce the main effect of each parameter: exponential suppression with increasing $\tilde{E}_J/E_C$ and sweet spots for $E_L/E_C$ and $\phi_0$. In Appendix \ref{appendix_harmonic_approx} we compare it with the numerical results, showing qualitative agreement. For the ASQ limit, the phase space is periodic and only one well is present, from which, the overlap becomes $\xi^{ASQ} \sim e^{-\frac{1}{2}\left(\frac{\pi -\phi_0}{\phi_{\rm zpf}}\right)^2}$ \cite{kurilovich2025andreevspinqubitprotected,Tjernshaugen2026coherentcontrolspinmons}.

\subsection{Anharmonicity}
\label{subsec:anharmonicity}
The above discussion demonstrates a good separation, {\it i.e.} small overlap, of spin-resolved ground state wavefunctions in phase space in the IPA regime. The gap to the excited states strongly depends on $E_L$, as seen by comparing Figs.~\ref{fig3}(b) and ~\ref{fig3}(c). This motivates the introduction of another important figure of merit: the anharmonicity, $\alpha$, defined as the energy distance between the qubit frequency $f_{01}$ and the next excited level $f_{12}$,
\begin{equation}
\alpha = (f_{12} - f_{01})=\varepsilon_{2}-2\varepsilon_{1} +\varepsilon_{0}.
\label{anharmon}
\end{equation}
The anharmonicity is a critical qubit design parameter, determining, for instance,
the minimum pulse duration $\tau\sim 1/|\alpha|$ needed to avoid leakage outside the computational states. For a nearly degenerate spin qubit subspace with $\varepsilon_{1} =\varepsilon_{0}$, the anharmonicity approaches $\alpha\approx \varepsilon_{2}-\varepsilon_1$ and, therefore $\tau\sim 1/\varepsilon_2$.

We focus in the IPA regime with $\delta\phi_{\rm ext}\approx 0$, Fig. \ref{fig3}, where only a small number of low-energy transitions determine $f_{01}$ and $f_{12}$, simplifying the discussion, see Fig.~\ref{fig2}(b). Indeed, at the time-reversal symmetry point $\delta\phi_{\rm ext}=0$, the potentials in each spin are mirrored in phase, providing a degenerate spectrum, {\it e.g.} Fig.~\ref{fig1}(b). The second transition $f_{12}=\alpha$ is also degenerate, with a spin conserving and a spin flipping line, see Fig.~\ref{fig2}(b).

Figure~\ref{fig3}(h) shows a contour plot of $\alpha$ as a function of $E_L/E_C$ and $\tilde{E}_J/E_C$, with line cuts in Fig.~\ref{fig3}(i). In the ASQ limit, $E_L=0$, the anharmonicity is governed by the plasma mode of the QD junction, given by  $\alpha^{ASQ}\simeq f_p=\sqrt{8E_C\tilde{E}_J}$~\cite{kurilovich2025andreevspinqubitprotected}, see arrow in Fig.~\ref{fig3}(i). Increasing $E_L$ raises the energy of the potential wells away from $\phi=0$. In this limit, $f_{12}$ is determined by transitions of fluxon type, $\alpha \approx t^f_{\sigma\sigma} \approx 2\pi\phi_0E_L/\chi$, approximation represented in Fig.~\ref{fig3}(i) with a red line. At zero flux, this transition coincides with the spin-flip transition between wells that are nearby, also described in Sec.~\ref{harmonic}, with green lines in Fig.~\ref{fig2}(b). Finally, in the inductive-dominated regime, $E_L\gg\tilde{E}_J$, the potential becomes approximately parabolic, {\it c.f.} Fig.~\ref{fig3}(d), the first excitation in each spin sector is of plasmon type, leading to $\alpha \approx f_p^L= \sqrt{8E_CE_L}$.

Overall, Figs.~\ref{fig3}(h,i) show that increasing $E_L$ enhances the anharmonicity. In the fluxonium case, the anharmonicity reaches a maximum for $E_L/E_C\sim 1-10$, depending on the Josephson energy. Further increasing $E_L$ makes the potential converge towards the harmonic potential. This harmonic limit is achieved for a dominating $E_L$ over the Josephson energy, leading to $\alpha\to0$. In contrast, $\alpha$ in the IPA case, the anharmonicity continues growing with $E_L$, even above the plasma energy $f_p$. In this respect, the IPA qubit can exhibit a larger $\alpha$ than the fluxonium qubit, see Fig.~\ref{fig3}(j). 

The preceding cases show that the intrinsic trade-off between phase-space separation and anharmonicity requires both quantities to be optimized simultaneously. Importantly, variations of the parameters in Eq.~\eqref{H-IPA} can change the overlaps and anharmonicity by several orders of magnitude, highlighting the possibility of optimizing different figures of merit depending on the device properties.

\begin{figure*}
    \centering
    \includegraphics[width=0.9\textwidth]{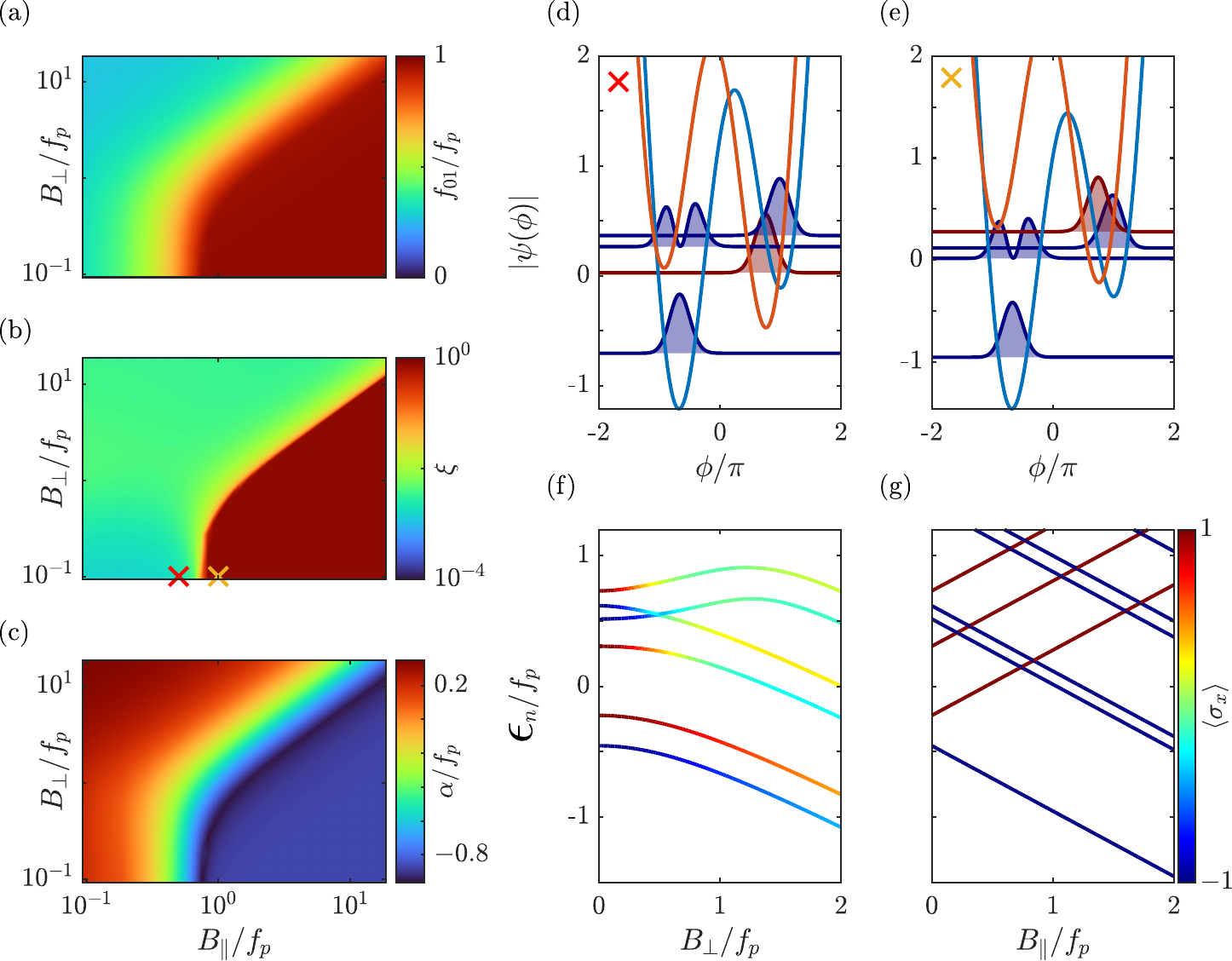}
\caption{(a) Qubit frequency ($f_{01}/f_p$), (b) wavefunction overlap ($\xi$), and (c) anharmonicity ($\alpha$) as a function of the magnetic field 
parallel/perpendicular to the SO axis, $B_\parallel$ and $B_\perp$, respectively. (d) Amplitudes of the lowest-energy states and spin potential branches normalized by plasma frequency $f_p$ for the situations marked in panel (b) with a red cross ($B_\perp/f_p=0$ and $B_\parallel/f_p=0.5$). (e) Same as (d) but for parameters corresponding to the orange cross in (b): $B_\perp/f_p=0$ and $B_\parallel/f_p=1$. (f) Eigenvalues evolution versus $B_\perp/f_p$ for fixed $B_\parallel/f_p=0$. (g) Eigenvalues evolution versus $B_\parallel/f_p$ for fixed $B_\perp/f_p=0$. The color indicates the spin projection along the SO axis $\langle\sigma_x\rangle$. 
The external flux is fixed to $\phi_{\rm{ext}}=0.05\pi$ so the ground state degeneracy is slightly broken at zero field. Rest of parameters: $E_L=5E_C$, $\tilde{E}_{J}/E_C=27.8$.}
       \label{fig5}
\end{figure*}

\begin{figure*}[t!]
    \centering
    \includegraphics[width=\linewidth]{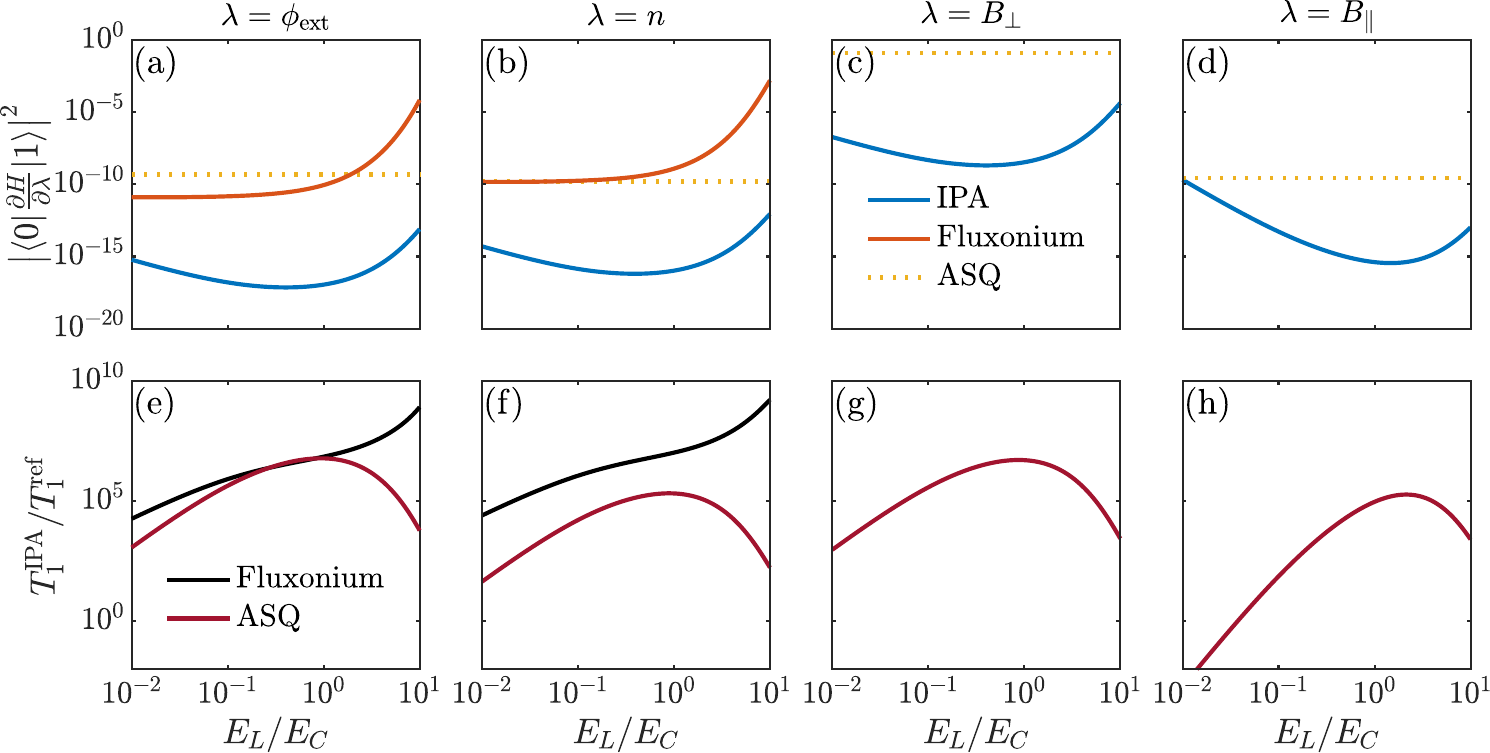}
    \caption{ Top panels: matrix element $|\langle0|\partial H/\partial \lambda|1\rangle|^2$ for $\lambda=\phi_{ext}$ (a), $\hat{n}$ (b), $B_z$ (c), and $B_x$ (d). The solid blue and orange curves present results for the IPA and the fluxonium, while the yellow dotted line represents the result for the ASQ. Bottom panels: ratio between the relaxation times, $T_1$, for the IPA compared to the fluxonium (black) and the ASQ (red), due to fluctuations in the same variables as in the top panels. Calculations are performed considering the different qubits feature the same $1/f$ noise spectral density, given in Eq.~\eqref{Eq:S_lambda}. Parameters: $E_0=2E_{\text{SO}}$, $\tilde{E}_{J}/E_c=27.8$, $B_z/E_c=3\cdot10^{-4}$, and $\delta\phi_{\text{ext}}=0.05\pi$. We consider $E_J=2E_0$ for the ASQ.}
    \label{fig:t1}
\end{figure*}

\subsection{Flux dependence}
\label{subsec:flux}
As mentioned before, the IPA features a double well potential at zero flux, $\phi^0_{\rm ext}=0$, similar to a heavy fluxonium biased at half-
flux, $\phi^0_{\rm ext}=\pi$. Shifting the external flux away from this condition $\delta\phi_{\rm ext}=\phi_{\rm ext}-\phi^0_{\rm ext}\neq 0$ breaks the time reversal symmetry, therfore introducing asymmetries between the two spin wells and lifting their energy degeneracy, Fig.~\ref{fig4}(b). This results in a linear increase on the qubit frequency with $\delta \phi_{ext}$. We illustrate this behavior in Fig. \ref{fig4}(d), where we benchmark the IPA case (solid lines) for increasing ratios $\tilde{E}_J/E_C$ against the corresponding fluxonium (dashed lines) and the harmonic approximation (dotted lines) from Eq. \eqref{qubit-freq-harmonic}.  For clarity, we only show this harmonic approximation for the least favorable case, namely for the lowest $\tilde{E}_J/E_C$, noting that agreement improves as $\tilde{E}_J/E_C$ increases and the ratio $E_L/\tilde{E}_J$ correspondingly decreases (not shown). 

The linear increase in qubit frequency is accompanied by a reduction in anharmonicity, which eventually crosses zero, Fig.~\ref{fig4}(e). This behavior arises from the decreasing energy difference between two states with the same spin located at different potential, see Fig.~\ref{fig4}(b). Eventually, these two states can cross, happening at $\phi_{ext}\sim 0.14\pi$, for the parameters used for the grey line in Fig.~\ref{fig4} (i.e $\tilde{E}_J/E_C=27.8$ and $E_L/E_C=5$). After the crossing, the two lowest-energy states have opposite spins and locate at nearby wells, Fig.~\ref{fig4}(c). The level crossing leads to a change on the qubit frequency and anharmonicity, Figs.~\ref{fig4}(d,e), and a sudden increase of the absolute overlap, Figs.~\ref{fig4}(f).

\subsection{Magnetic field effects\label{magnetic field}}
In the minimal description, excluding a possible phase-dependent spin splitting due to a variant of the Knight shift~\cite{SciPostPhys.15.2.070}, an external magnetic field leads to a Zeeman splitting that can be described through the term
\begin{equation}
\label{Zeeman}
H_Z=\vec{B}\cdot\vec{\sigma},
\end{equation}
added to Eq.~\eqref{QD-JJ}.
This Zeeman term leads to two modified Josephson branches in the quantum dot potential as

\begin{equation}
\label{spin-branchesEz}
U_{\pm}^{QD}({{\phi}})=E_0\cos({{\phi}}) \pm \sqrt{B_\perp^2+(B_{||}+E_{SO}\sin{({\phi})})^2},
\end{equation}
such that a Zeeman field $B_{||}$
parallel to the SO axis shifts the energy of the spin branches while an orthogonal field $B_\perp$, 
opens avoided crossings at degeneracy points in phase space \cite{Bargerbos2023b}.

Increasing $B_{||}$ lifts the up spin branch relative to the down spin, see red and blue lines in Figs.~\ref{fig5} (d,e). Since the IPA uses the two lowest-energy states with opposite spins along the SO direction, increasing $B_{||}$ leads to a linear increase of the qubit frequency, as illustrated in Fig.~\ref{fig5}(a) [see also Fig.~\ref{fig5}(g)]. Eventually, $f_{01}$, saturates to $f_p$ for sufficiently large $B_{||}$ values, indicating a level crossing between the lowest-energy and the first excited state in the up and down spin wells, illustrated in Figs.~\ref{fig5}(d,e).

As seen in Fig.~\ref{fig4}, the crossing between the two spin states, in this case due to magnetic fields, is also reflected as an abrupt change in the wavefunction overlap, which grows from a very suppressed value, $\xi\ll1$, to $\xi\sim1$ as shown in Fig.~\ref{fig5}(b), and the qubit anharmonicity, which goes from positive to negative values as $f_{01}$ becomes larger than $f_{12}$.  These results illustrate the transition from two well-isolated spins to a qubit confined in a single well potential.

On the other hand, $B_\perp$ hybridizes the two spin states, {\it c.f.} Eq.~\eqref{spin-branchesEz}, as illustrated by the color of the lines in Fig.~\ref{fig5}(f). This hybridization enlarges the region with small wavefunction overlap, see Fig.~\ref{fig5}(b).

An alternative approach to understanding the effect of moderate magnetic fields is to project $H_Z$ onto the lower energy states of the circuit, making use of the harmonic approximation states $\ket{\psi_{\sigma}^{m,n}(\phi)}$, see Eq.~\eqref{eq:general states wf}. 
While the parallel field shifts the states in the direction of their spin, $\langle\psi_{\sigma}^{m,n}(\phi)|B_\parallel\sigma_\parallel|\psi_{\sigma'}^{m',n'}(\phi)\rangle=\delta_{\sigma\sigma'}\delta_{nn'}\sigma_{\parallel} B_\parallel$,  the perpendicular part couples states with opposite spin 
$\langle\psi_{\sigma}^{m,n}(\phi)|\vec{B}_\perp\vec{\sigma}_\perp|\psi_{\sigma'}^{m,n'}(\phi)\rangle=S_{m\sigma n,m'\sigma'n'}B_\perp \bra*{\sigma}\vec{\sigma}_\perp\ket*{\sigma'}\propto \delta_{\sigma\bar{\sigma}'} \delta_{n{n}'}$.
This coupling is reduced by the overlap between the bare wavefunctions   $S_{m\sigma n,m'\sigma'n'} = \int d\phi\,  \psi_{\sigma}^{m,n}(\phi)\psi_{\sigma'}^{m',n'}(\phi)$. In the regime where the harmonic approximation applies (Sec. \ref{harmonic}), the overlaps exponentially decrease with the distance between the corresponding wells, so the coupling is dominated by the wavefunctions closest in phase, even if they are at higher energies; see also \cite{kurilovich2025andreevspinqubitprotected}.

\subsection{Qubit operation}
\label{subsec:qubitOperations}

Qubit manipulations can be achieved via multiple routes: driving the magnetic field 
$B_{\perp}$; modulating the gate voltage $V_g$, which affects both 
$E_{0}(V_g)$ and $E_{SO}(V_g)$ and enables electric dipole spin resonance (EDSR)~\cite{Nadj-Perge2010,PitaVidal2023} by coupling the charge to an oscillating electric field; or varying the external flux $\delta\phi_{\rm ext}$, which tunes the superconducting phase and thereby the effective spin–orbit interaction. Each of these control knobs provides a distinct pathway for qubit operations, allowing one to choose the most robust method against specific decoherence sources (see section \ref{decoherence-section}) or to combine them for more complex manipulation sequences.

To understand how $V_g$ and $\delta\phi_{\rm ext}$ can induce spin transitions, we can revisit the situation of ASQs driven by gate~\cite{Tosi2019,Metzger2021,Bargerbos2023b,PitaVidal2023,Lu2025} or by flux~\cite{Hays2020,Hays2021}. Such a driving is possible because the states have a \textit{pseudospin} whose orientation depends on those parameters. The minimal model of the junction in Eq.~\eqref{QD-JJ} requires a generalization to include a $V_g$ and $\phi$ dependence of the SO direction, $\vec{n}_{\rm SO}$ in Eq.~\eqref{QD-JJ}, as noted in Ref.~\cite{kurilovich2025andreevspinqubitprotected}. The spin flipping transitions thus require finite matrix elements for $\partial \vec{n}_{SO,\perp}/\partial V_g$ and $\partial \vec{n}_{SO,\perp}/\partial \phi$, with the unit vector $\vec{n}_{SO}=\vec{E}_{SO}/|\vec{E}_{SO}|$, being $\vec{n}_{SO,\perp}$ the unit vector which is perpendicular to $\vec{n}_{SO}$  at fixed $V_g$ and $\delta\phi_{ext}$ (note that the flux driving can provide spin flipping only if part of the phase drop $\delta\phi_{ext}$ occurs in the weak link \cite{You2019,caceres2026}). From the gate driven experiments, which probe the direct spin flip transition, we can confirm the required dependence of $E_0$ and $E_{SO}$ on $V_g$. 

As discussed in Ref.~\cite{kurilovich2025andreevspinqubitprotected}, higher energy levels are more extended in phase space and can significantly feature a finite overlap with states located at different Josephson potential wells. It is possible to induce transitions between both ground states, via Raman pulses through higher energy levels that never get populated. This fact may be used to manipulate the qubit without the need of a Zeeman field.

A bit out of the degenerate qubit situations, $\delta\phi_{ext}\neq0,\pi$ (mod $2\pi$), we can have the \textit{asymmetric structure} of Fig.~\ref{fig2}(c) in the energy spectrum required for the Raman protocol, while maintaining the disjointness between the ground states. Using an excited intermediate state with different spin, an almost resonant drive can induce a spin-flip transitions. For instance, taking $\psi_{+}^{0,0}(\phi)$ (qubit state with $\sigma=+$ spin) and $\psi_{-}^{1,0}(\phi)$ (qubit state with $\sigma=-$ spin), using the intermediate state $\psi_{-}^{0,0}(\phi)$, we can induce transition from the qubit spin $\sigma=+$ state
via spin-flipping driving at $t_{+-}^p$. Note that $\psi_{+}^{0,1}(\phi)$ is not populated even if the drive also has a spin-conserving component, because of the frequency mismatch with the Raman pulse. 

On the other hand, to induce transitions between $\psi_{-}^{0,0}(\phi)$ and $\psi_{-}^{1,0}(\phi)$, we can perform a spin-conserving drive at  $t_{--}^f$, while maintaining  $\psi_{+}^{0,0}(\phi)$ unpopulated. In particular, if $t_{--}^p(k) \neq t_{++}^p(k)$, for any $k$ being an upper energy state transition, which is possible in this loop configuration, we can use both drivings in a STIRAP protocol that avoids the need of a sequential process populating the intermediate state --- similarly to what has been implemented in ASQs~\cite{Hays2021,Cerrillo2021} \footnote{In \cite{Cerrillo2021}, the intermediate states are other ABS manifold, with which the lower one has a quite larger spin flipping matrix element (this is not our situation, with presumably weak spin flipping efficiency).}

To perform the opposite transfer, one can apply the reverse operation, or the reciprocal one with $\psi_{+}^{1,0}(\phi)$ as the intermediate state. It is convenient to use the lowest-possible $n$ that extends over the two wells, as increasing $n$ leads to states that resemble the ones of an LC resonator, where $t_{--}^p(k)\approx t_{++}^p(k)$. The energy distance of the intermediate state to neighbouring levels is also relevant to allow a fast operation that does not populate them. It will depend on the regime of the IPA, {\it e.g.} a large \textit{kinetic} term may already rise $n=1$ out of its well; in any case, there are generalizations of the STIRAP protocol that withstand those neighboring levels~\cite{Cerrillo2021}.

\section{Relaxation and noise susceptibilities \label{decoherence-section}} 
Protection against local noise often involves a tradeoff between conjugate quantum variables. This behavior is particularly transparent in superconducting circuits, where the superconducting phase ($\phi$) and Cooper-pair number ($n$) form a conjugate pair. For example, the transmon is designed to exponentially suppress its sensitivity to low-frequency charge noise by operating at large ($E_J/E_C$). In this regime, however, the transition matrix element ($\left|\langle0|\hat n|1\rangle\right|$) remains large, allowing noise to induce energy relaxation. Conversely, fluxonium qubits can exhibit strongly suppressed charge matrix elements and correspondingly weak relaxation through capacitive or dielectric channels \cite{PhysRevX.9.041041,PhysRevX.11.011010}, while retaining appreciable sensitivity to flux fluctuations away from flux sweet spots. Thus, engineering protection against one class of local perturbations can enhance sensitivity to another.

In what follows, we discuss the IPA qubit response against different sources of noise, including magnetic and flux noise, and benchmark it against the ASQ and the fluxonium. Hereafter, we use $\ket{0,1}$ as a short hand notation for the lowest-energy wavefunctions, $\psi_{0,1}(\phi)$.

\begin{figure}
    \centering
\includegraphics[width=\columnwidth]{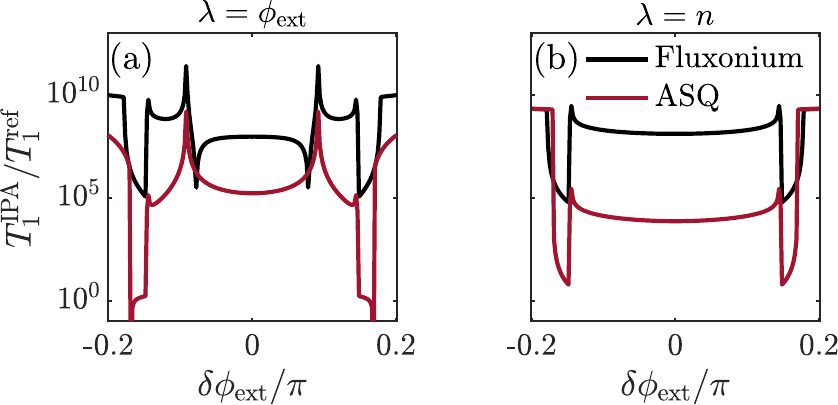}
\caption{Ratio between the relaxation times, $T_1$, for the IPA compared to the fluxonium (black) and the ASQ (red) for flux noise (panel a) and charge noise (panel b) against $\delta\phi_{\text{ext}}$. Parameters: $E_L=5E_C$, $E_0=2E_{\text{SO}}$, $\tilde{E}_{J}/E_c=27.8$, and $B_z/E_C=3\cdot10^{-4}$. We consider $E_J=2E_0$ for the ASQ.}
\label{fig11}
\end{figure}
\subsection{Relaxation rates and times \label{relaxation-decoherence}}
To determine the relaxation rates, we use the Bloch-Redfield approximation~\cite{Krantz_IOP2019}
\begin{equation}
    \Gamma_1(\lambda)\equiv\frac{1}{T_1(\lambda)}= 4\pi^2\left| \left\langle 0 \left| D_\lambda \right| 1 \right\rangle \right|^2 \cdot S_{\lambda}(f_{01})\,,
    \label{eq:relaxRate}
\end{equation}
where $ D_\lambda=\frac{\partial H}{\partial \lambda}$ is the noise operator associated with a given noise source that affects the parameter $\lambda$. As the IPA qubit frequency is typically small, we consider $1/f$ noise 
\begin{equation}
    S_\lambda(f_{01})=\frac{A}{f_{01}}\,,
    \label{Eq:S_lambda}
\end{equation}
where $A$ is a non-universal constant that is system-dependent and $f_{01}$ is the qubit frequency.

For phase and charge fluctuations,
\begin{equation}
\label{noise-operator1}
\begin{aligned}
& D_{\phi_{\mathrm{ext}}}^{\mathrm{IPA/ASQ}} = E_0 \sin(\phi - \phi_{\mathrm{ext}}) \hat{\sigma}_0 + E_{\mathrm{SO}} \cos(\phi - \phi_{\mathrm{ext}}) \hat{\sigma}_x \,, \\
&D_{\phi_{\mathrm{ext}}}^{\mathrm{Flux}} = -E_J \sin(\phi - \phi_{\mathrm{ext}}) \hat{\sigma}_0 \,.\\
& D_{n} = 8E_C \hat{n} = 8E_C(-i\partial_\phi).
\end{aligned}
\end{equation}
The flux noise operator incorporates a standard noise contribution from the Josephson coupling $E_0$ in addition to a novel, spin-dependent, term proportional to $E_{\text{SO}}$. Both terms could also exhibit fluctuations induced by gate noise from mesoscopic fluctuations of the QD parameters $E_0(V_g)$ and $E_{\text{SO}}(V_g)$ that are not considered here. The second equation in Eq.~\eqref{noise-operator1} constitutes a standard charge noise term in superconducting qubits, describing, for example, energy relaxation arising from two-level fluctuations in the capacitor dielectrics.

In addition to the noise terms typical of superconducting circuits, magnetic noise is the dominant decoherence source in spin-based qubits. Consequently, it must be included when comparing the IPA with the ASQ. The Zeeman term in Eq.~\eqref{Zeeman} not only describes the intentionally applied magnetic field used to manipulate the spin qubit but also accounts for random magnetic fields arising from the surrounding environment. Without delving into a microscopic derivation, which could involve, for example, hyperfine nuclear noise \cite{hyperfine2005,Reilly_2008}, we consider a magnetic noise term, which adds to the Zeeman term, of the form
\begin{equation}
H_m=\delta B_{||}\sigma_x+\delta B_{\perp}\sigma_z.   
\end{equation}
Therefore, the two terms describing such magnetic noise are just Pauli matrices parallel and perpendicular to the SO axis, respectively.

Results are shown in Fig.~\ref{fig:t1}, where the four columns indicate the four main source of errors: flux fluctuations (first column), charge fluctuations, (second column), and perpendicular and parallel magnetic fields to the spin-orbit direction (third and fourth columns, respectively. Flux noise is the dominant source of error of fluxonium, while magnetic field fluctuations limit the performance of ASQs.

We begin our analysis discussing the dominant relaxation mechanisms. The top panels in Fig.~\ref{fig:t1} display the modulus of the matrix element in Eq.~\eqref{eq:relaxRate} as a function of $E_L$, that quantifies how fluctuations in a given parameter connect the two qubit states, therefore determining the relaxation rate. The matrix elements feature a minimum for $E_L\gtrsim E_c$, being several orders of magnitude lower than the marix element for the ASQ and the fluxonium qubit, highlighting a significantly increased protection against relaxation.

The improved protection against relaxation is better illustrated in Figs.~\ref{fig:t1}(e-h), where we display the ratio between the expected $T_1(\lambda)$ for the IPA and the fluxonium (black line) and the ASQ (red line). To make the comparison general, we have considered that the noise distribution for the three qubits, $S_\lambda(f_{01})$, is the same, {\it i.e.}, the constant prefactor $A$ in Eq.~\eqref{Eq:S_lambda} is the same for the three cases.

Numerical calculations, shown in the lower panels in Fig.~\ref{fig:t1}, indicate an increase on $T_1$ times of about five orders of magnitude compared to the ASQ and the fluxonium at the optimal point. This highlights the improved protection of the IPA, arising from the interplay between the spin-resolved Josephson potential and the parabolic confinement.

Finally, we show the ratio between the $T_1$ times for flux and charge fluctuations in Fig.~\ref{fig11} as a function of $\delta \phi_{\rm ext}$. In both cases, the IPA features an enhanced $T_1$ with respect to the fluxonium and the ASQ for a region close to $\delta \phi_{\rm ext}\sim0$. 
On the other hand, there is a level crossing between the $|1\rangle$ qubit state and the first excited one for  $|\delta \phi_{\rm ext}|\approx 0.14\pi$ for the chosen parameters, see also Fig.~\ref{fig4}, that leads to a jump on the ratio.

\subsection{Dephasing}
We now analyze the effect of fluctuations on the dephasing times, $T_\phi$. Here, we use a low-frequency approximation, consistent with the $1/f$ noise considered in Eq.~\eqref{Eq:S_lambda}. In this limit, the dephasing rate, as measured in a Ramsey experiment, 
can be approximated to lowest order
as~\cite{Ithier_PRB2005,Krantz_IOP2019,Geier2024}, 
\begin{equation}
\label{T_phi_def}
    \Gamma_\phi(\lambda) \equiv\frac{1}{T_\phi(\lambda)} \approx A\,c_1\left( \frac{\partial f_{01}}{\partial \lambda} \right),
\end{equation}
which is valid outside sweet spots. Here, $c_1$ is a constant depending on the low frequency cutoff of the $1/f$ noise and the measurement time. In what follows we compare the dephasing of the three qubits discussed in the paper which, asuming similar experimental noise and measurement conditions, can be quantified by just the ratio of derivatives of the qubit frequencies.
\begin{figure}[t!]
    \centering   \includegraphics[width=0.50\textwidth]{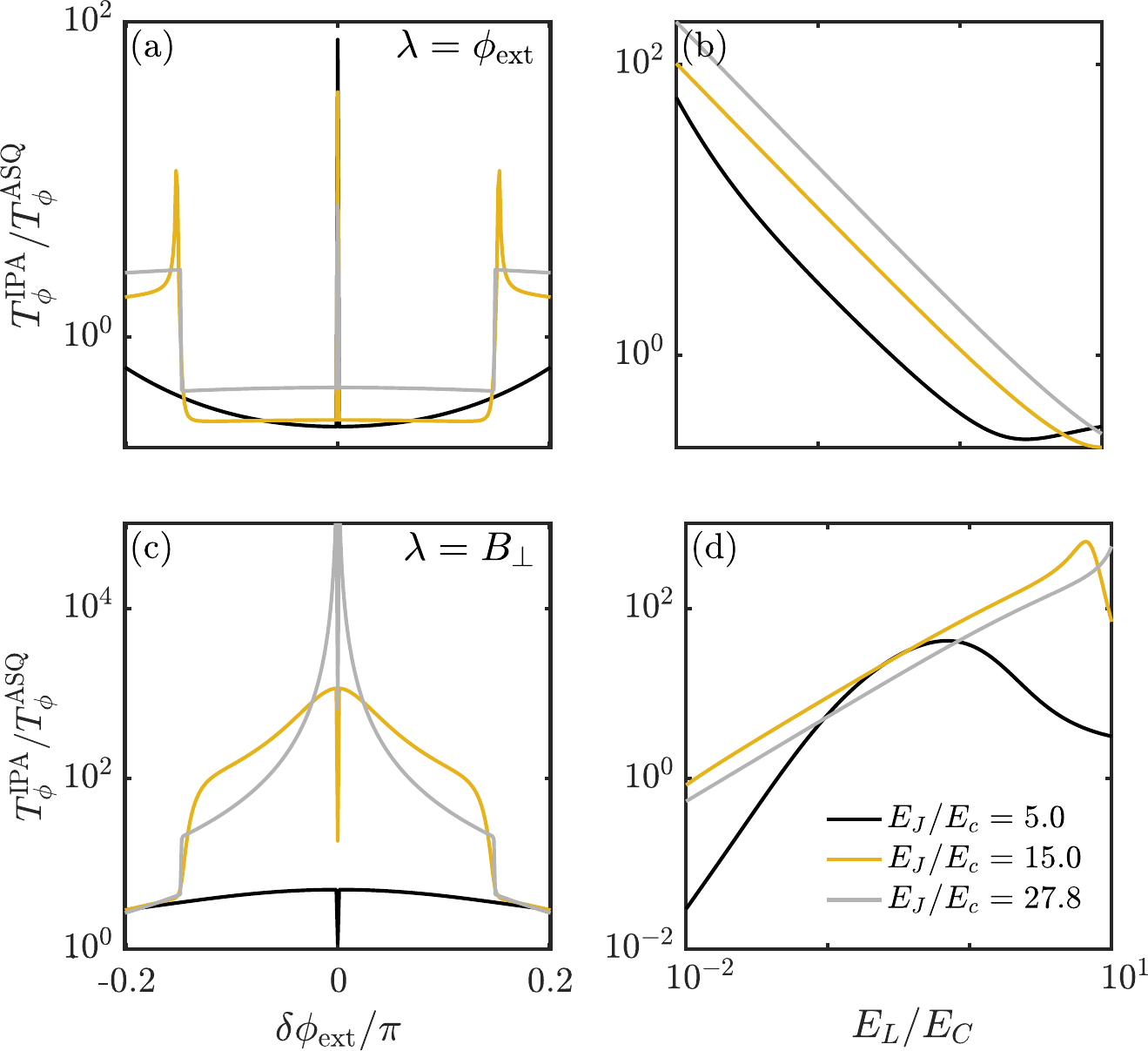}
 \caption{Ratio of the $T_{\phi}$ times for the IPA and the ASQ, see Eq.~\eqref{T_phi_def}, as a function of flux $\delta \phi_{\rm{ext}}$ (left panels) and $E_L$ (right panels), for different values of  $\tilde{E}_J/E_c$: 5 (black), 15 (gold), and 27.8 (grey). Panels (a,b) present results for flux fluctuations, while panels (c,d) show the effect of magnetic field fluctuations perpendicular to the SO direction. Fluctuations along the SO, $\lambda=B_x$, lead to $T^{\rm IPA}_\phi(B_x)/T^{\rm ASQ}_\phi(B_x)\approx1$ (not shown). We use $E_L/E_C=5$ for the left panel and $\delta \phi_{\rm ext}=0.05\pi$ for the right ones.}
    \label{fig7}
\end{figure}

The ratio between the dephasing times, $T_\phi$, between the IPA and the ASQ is shown in Fig.~\ref{fig7} for flux (top panels) and magnetic fluctuations perpendicular to the spin-orbit direction (bottom panels). Different lines correspond to different values of $\tilde{E}_J$, same cases as analyzed in Fig.~\ref{fig3}. For values around the $\delta\phi_{\rm ext}=0$, the ratio between the dephasing times is flat, signature of the linear increase of the qubit frequency with flux for both, IPA and ASQ. The sharp feature at $\delta\phi_{\rm ext}=0$ originates from the vanishing first-order flux sensitivity, $\partial f_{01}/\partial \phi_{\rm ext}=0$. At this sweet spot, however, the first-order approximation in Eq.~\eqref{T_phi_def} is no longer applicable, and dephasing is instead determined by higher-order flux sensitivity. Moreover, the qubit is nearly degenerate at this point, making operation challenging. Finally, The sudden increase on the $T_\phi$ for largest $\delta \phi_{\rm ext}$ is due to a level crossing between the qubit $|1\rangle$ state and the first excited state, see Fig.~\ref{fig4}. 

In the IPA regime, $E_L\gtrsim E_c$, the IPA qubit is more susceptible to flux noise than the ASQ. Decreasing $E_L$ reverses this trend. As illustrated in Fig.~\ref{fig7}(b), decreasing $E_L$ leads to a monotonous increase on $T_\phi(\phi_{\rm ext})$, overcoming the value obtained for the ASQ by a few orders of magnitude. This is due to the delocalization of the qubit wavefunctions along several minima of the Josephson potential, similar to what happens in the Blochnium regime \cite{Blochnium2020}.

As already mentioned, magnetic noise is the leading dephasing source of noise in spin-based qubits. In Figs.~\ref{fig7}(c,d) we show the ratio between the $T_\phi$ calculated for a magnetic noise in the $z$ direction, perpendicular to the SO direction. As shown, the Josephson profile of the IPA qubit enhances the protection against magnetic field noise. This is true for a range of $E_L$ and $\tilde{E}_J$ values, see Fig.~\ref{fig7}(d). In contrast to the flux noise case, increasing $E_L$ initially leads to an increase of $T_\phi(B_z)$. This is due to the improved localization of the spin wavefunction in the two wells that decreases the energy dependence on the Zeeman field, $B_z$. For large $E_L$ values, the Josephson potential tends to the harmonic potential, see Fig.~\ref{fig4}(d), increasing the hybridization between the qubit states induced by $B_z$. Finally, we note that $T^{IPA}_\phi(B_x)/T^{ASQ}_\phi(B_x)=1$ close to $\delta\phi_{\rm ext}\sim 0$ for a large range of $E_L$ values (not shown), denoting no improved protection with respect to spin fluctuations along the SO direction.

\begin{figure}[t!]
    \centering    \includegraphics[width=0.5\textwidth]{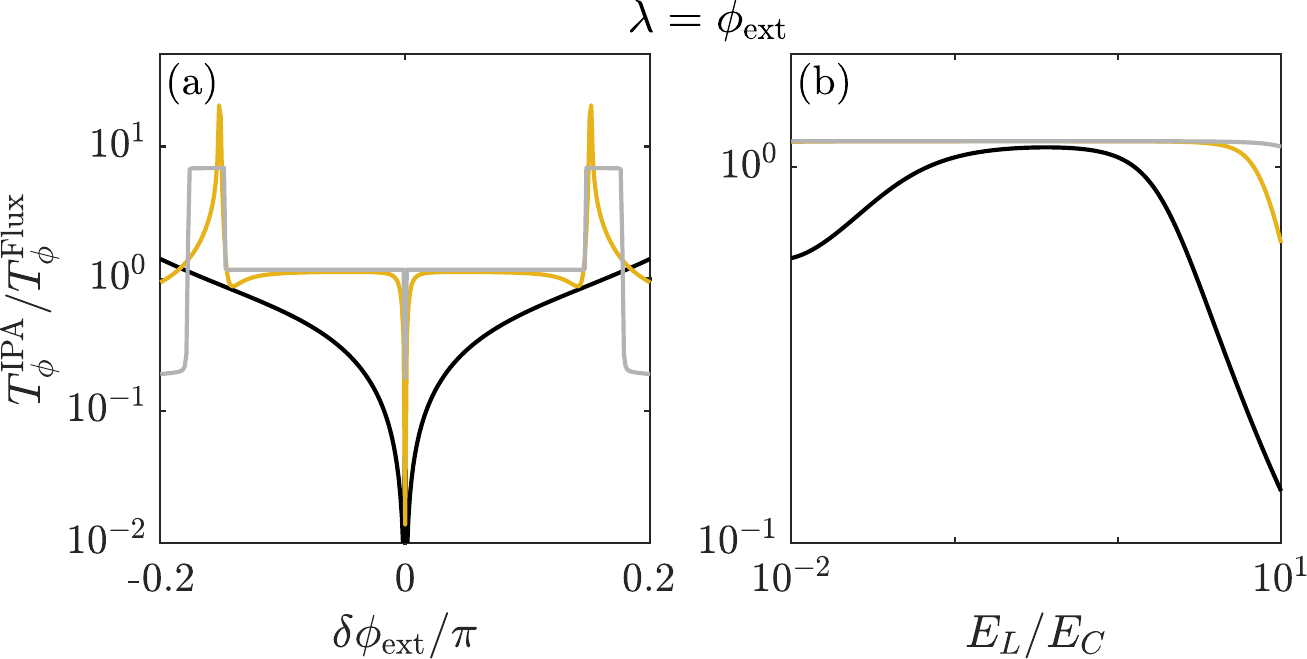}
    \caption {$T_{\phi}$ ratios of IPA divided by Fluxonium as a function of (a) flux and (b) $E_L$ for $\tilde{E}_J/E_C$= 5 (black), 15 (gold), and 27.8 (grey) respectively. $E_L/E_C=5$ in panel (a) and $\delta \phi_{\rm ext}=0.05\pi$ in panel (b).}
    \label{fig8}
\end{figure}

We conclude by comparing $T_\phi$ for the IPA and the fluxonium, see Fig.~\ref{fig8}. For flux noise, dominant dephasing mechanism of the fluxonium, the $T_\phi$ features a plateau with a value close to one near the flux sweet spot. We note that this plateau corresponds to the regime where the IPA featured increased $T_1$ protection, see Fig~\ref{fig11}(a) for comparison. The two peaks with enhanced flux noise protection at $\delta\phi_{\rm ext}\approx\pm0.14\pi$ correspond to the crossing between the qubit $|1\rangle$, that locates at a different well than the qubit $|0\rangle$ state, with the first excited state, that locates in the same well as the $|0\rangle$ state, {\it i.e.}, from fluxon to plasmon transitions.

The case with the smallest Josephson energy, black line in fig.~\ref{fig8}, shows features that are smoother for the fluxonium, due to the enhanced hybridization between the states in the different wells. This results on a decrease on the ratio of $T_\phi$ between the IPA and fluxonium for a broad range of $\delta \phi_{\rm ext}$.

The comparison between the IPA and the fluxonium is better illustrated in Fig.~\ref{fig8}(b), where we show the ratio between their respective $T_\phi$ for a small $\delta \phi_{\rm ext}\neq 0$. The comparison shows that both qubits have a similar $T_\phi(\phi_{\rm ext})$ performance for a broad range of $E_L/E_C$ values. For small ($E_L\ll E_C$) and large ($E_L\gg \tilde{E}_J$)  values, the tunneling between the different wells in the fluxonium case leads to a wavefunction that lives in both of its potential minima, increasing the flux resilience. On the other hand, the tunneling between the different minima is further suppressed in the IPA case, due to their spin-resolved character, resulting on a lower $T_\phi({\rm \phi_{ext}})$ in these two limits. Interestingly, in the protected regime, where Josephson energy dominates and $E_L/E_C\sim1$, the IPA can feature a slight increase on the dephasing times compared to the fluxonium. Conversely, smaller values of the Josephson energy lead to better fluxonium dephasing times with $T_\phi$ ratios of the order $10^{-1}$ as one approaches $E_L>E_C$.
 
\section{Conclusions and outlook}
\label{Sec:conclusions}
In this work, we have introduced and theoretically analyzed the inductively protected Andreev (IPA) spin qubit, a hybrid architecture that addresses the critical challenge of decoherence in Andreev spin qubits (ASQs). By integrating a semiconductor quantum dot (QD) Josephson junction into a fluxonium-like circuit, we have demonstrated that the addition of a linear inductor fundamentally alters the qubit's energy landscape, conferring significant protection against bit-flip processes and flux noise sources while being gate and flux tunable.

The experimental performance of the IPA qubit depends on the choice of semiconductor material. Our analysis suggests that spin fluctuations are the primary dephasing mechanism for both the ASQ and the IPA, which naturally directs attention toward material platforms that minimize such noise. While early experiments on ASQs have successfully been implemented in InAs nanowires~\cite{Hays2021,PitaVidal2023} due to their strong spin–orbit interaction, these systems are susceptible to hyperfine noise arising from nuclear spins. This intrinsic source of magnetic noise fundamentally limits the achievable coherence times and underscores the need for materials with a reduced nuclear spin bath. In this context, proximitized germanium (Ge ) \cite{PinoPRB2025,BabkinPRB2025,fabris2026granularaluminuminducedsuperconductivity} has emerged as a leading hybrid semiconductor–superconductor platform 
\cite{pitavidal2025novelqubitshybridsemiconductorsuperconductor} for Andreev qubit realizations \cite{hoffman2025resolvingandreevspinqubits,coppini2026strainengineeringandreevspin}. Germanium offers two decisive advantages: first, it can be isotopically purified to remove the small fraction of nuclear spins that would otherwise cause decoherence; second, its compatibility with established silicon-based fabrication techniques promises a path toward large-scale, high-yield device integration. Furthermore, the spin orbit interaction in Ge, which can be enhanced via strain engineering \cite{coppini2026strainengineeringandreevspin}, is essential for achieving the spin-resolved phase shifts that underpin the IPA mechanism. Therefore, Ge-based heterostructures, combined with high-kinetic-inductance superconducting elements, represent a highly promising materials platform for the first experimental demonstrations of the IPA qubit. Progress towards this goal includes the recent experimental demonstration of good proximity effect of Ge quantum wells proximitized by granular aluminium (grAl) \cite{fabris2026granularaluminuminducedsuperconductivity}. The large induced gap, combined with
grAl's large kinetic inductance and its magnetic-field resilience, establishes this hybrid as a versatile platform for exploring the ideas discussed here.

Looking ahead, several avenues warrant further investigation. While we have focused on protection against magnetic and flux noise, a detailed microscopic analysis of other decoherence channels affecting $E_0$ or $E_{SO}$ directly will be essential for a complete assessment of the IPA's performance. Furthermore, the scalability of this platform is promising. The spin-resolved nature of the qubit could enable novel schemes for long-range spin-spin coupling mediated by the superconducting circuit \cite{Pita-VidalNP2024}, and the ability to operate in a regime largely insensitive to external noise could relax constraints on fabrication and control electronics. Finally, exploring the interplay between the IPA design and other protected qubit architectures, such as the $\cos 2{\phi}$ or the $0-\pi$ qubits, may reveal new opportunities for hybrid quantum information processing.

In summary, we have shown that the IPA architecture offers a promising route toward enhancing the coherence of ASQs by leveraging the complementary strengths of semiconductor spin physics and protected superconducting circuits, while preserving their tunability and fast operation. The predicted orders-of-magnitude improvements in relaxation times, combined with tentatively competitive dephasing and large anharmonicity, establish the IPA as a compelling candidate for future high-coherence qubit implementations. Given the rapid experimental progress in hybrid semiconductor–superconductor platforms \cite{pitavidal2025novelqubitshybridsemiconductorsuperconductor}, including grAl induced superconductivity in Ge quantum wells \cite{fabris2026granularaluminuminducedsuperconductivity}, we expect that the key ingredients of the IPA design are within reach of current fabrication capabilities, following the first demonstrations of hybrid fluxonium qubits \cite{PhysRevApplied.14.064038,PRXQuantum.6.010326,isakov2026}. 

\section{Acknowledgements}
\noindent 
We acknowledge fruitful discussions with G. Steffensen and A. Maiani. We acknowledge funding from the
the Spanish Ministry of Science, Innovation, and Universities through AEI Grants CEX2024-001445-S (Severo Ochoa Centres of Excellence program), PID2024-161156NB-I00 and PID2022-140552NA-I00, and the Spanish Comunidad de Madrid (CM) “Talento Program”
(Project No. 2022-T1/IND-24070).
\appendix

\appendix

\section{Overlaps within the harmonic approximation} \label{appendix_harmonic_approx}

To model the lowest-energy states of the spin-dependent Josephson potentials, we applied the harmonic approximation for a single well in section \ref{harmonic}, which remains accurate in the regime $\tilde{E}_J\gg E_C \sim E_L$. 

In the regime when two fluxon wells are occupied in each spin sector ($B=0$), we can qualitatively describe the lowest energy  states as the harmonic ground states in each well, {\it c.f.} Eq. \eqref{harmonic-well}, coupled through phase slip processes

\begin{equation}
H = \begin{pmatrix}
    U_{\sigma}^m & E_S \\
    E_S &  U_{\sigma}^{m'}
    \label{Eq:H_harmonic_aapendix}
\end{pmatrix}.
\end{equation}

The fluxon tunneling rate $E_S$ can be approximated in the limit $\tilde{E}_J/E_C \gg 1$ as:
\begin{equation}
\label{eq:fluxon_tunnel}
E_S = A_S e^{-\sqrt{8\tilde{E}_J/E_C}} \quad \text{with} \quad A_S = 4 \left( \frac{8\tilde{E}_J^3 E_C}{\pi^2} \right)^{1/4}.
\end{equation}
The model in Eq.~\eqref{Eq:H_harmonic_aapendix} has two eigenvalues, $\lambda_{\pm} = \bar{V}_{\sigma} \pm \sqrt{h^2_{\sigma} + E_S^2}$, with
$\bar{V}_{\sigma}=(U_{\sigma}^m + U_{\sigma}^{m'} )/2$ and detuning $h_{\sigma}=(U_{\sigma}^m - U_{\sigma}^{m'}) / 2$. The hybridization between the two fluxons of the same spin is encoded in a tangent function as

\begin{equation}
\label{theta}
\tan(\theta_\sigma) = \frac{E_S}{h_{\sigma}} =\frac{E_S}{ \pi E_L(\delta \phi_{\rm ext}+\sigma\phi_0)/\chi}
\end{equation}

The ground state of each spin $\sigma$ is a linear combination of two fluxons $m=0$ and $m=1$ as
\begin{equation}
\label{eq:ground states}
\ket{g,\sigma} \approx  \sin (\theta_\sigma/2) \ket{m=0,\sigma} - \cos (\theta_\sigma/2) \ket{m=1,\sigma},
\end{equation}
which, using Eq. \eqref{eq:general states wf}, can written in phase space as
\begin{equation}
\label{eq:phase_space_wf}
\psi_{g\sigma}(\phi) = \frac{\psi_{\sigma}^{0,0}(\phi)\sin\left(\frac{\theta_\sigma}{2}\right) - \psi_{\sigma}^{1,0}(\phi)\cos\left(\frac{\theta_\sigma}{2}\right)}{\sqrt{1-\sin(\theta_{\sigma})S_{\sigma}}}
\end{equation}
\begin{widetext}
with $S_{\sigma}=\int d\phi \, |\psi_{\sigma}^{0,0}(\phi)| |\psi_{\sigma}^{1,0}(\phi)|=e^{-\frac{\kappa}{2} \frac{(\varphi_{\sigma}^0 -\varphi_{\sigma}^1)^2}{\chi^2} }=e^{- \frac{\pi^2}{2\chi^{3/2}\phi_{\rm zpf}^2}}$, such that $S_{+}=S_{-}$.
Using the GS wavefunctions in \eqref{eq:phase_space_wf} we can compute the overlap 

\begin{gather}
\label{eq:overlap^h}
\xi^{h} = \frac{1}{N(\theta)}\int d\phi \, |\psi_{g+}| |\psi_{g-}|
=  \frac{1}{N(\theta)}\left|\sin(\theta)S_{0+0,0-0} - \sin^2\left(\frac{\theta}{2}\right)S_{0+0,1-0} - \cos^2\left(\frac{\theta}{2}\right)S_{0-0,1+0}\right|, 
\end{gather}
where $N(\theta)$ is the wavefunction normalization and we have defined $S_{m\sigma,n,m'\sigma',n'}=\int d\phi |\psi_{\sigma}^{m,n}| |\psi_{\sigma'}^{m',n'}|$, $N(\theta)=1-S\cdot \sin(\theta)$, with $S=S_+=S_-$, and angles $\theta_+ =\theta$ and $\theta_- =\pi-\theta$.
\end{widetext}
Using the wave functions in Eq. \eqref{eq:general states wf}, the explicit expressions of the overlaps entering Eq. \eqref{eq:overlap^h} read

\begin{align}
\label{eq:same_fluxon_overlap}
    S_{0+0,0-0} &= e^{-\kappa(\varphi_{+}^0-\varphi_{-}^0)^2} = e^{\left( \frac{-1}{2\chi^{3/2}} \left(\frac{\phi_0}{\phi_{\rm zpf}}\right)^2 \right)}=S_{1+0,1-0} \\
    \label{eq:xi}
    S_{0+0,1-0} &= e^{-\kappa(\varphi_{+}^0-\varphi_{-}^1)^2} = e^{\left( \frac{-1}{2\chi^{3/2}} \left(\frac{\pi -\phi_0}{\phi_{\rm zpf}}\right)^2 \right)} \\
    \label{eq:xi*}
    S_{0-0,1+0} &= e^{-\kappa(\varphi_{-}^0-\varphi_{+}^1)^2} = e^{\left( \frac{-1}{2\chi^{3/2}} \left(\frac{\pi +\phi_0}{\phi_{\rm zpf}}\right)^2 \right)}
\end{align}

\begin{figure}[t!]
    \centering
    \includegraphics[width=\columnwidth]{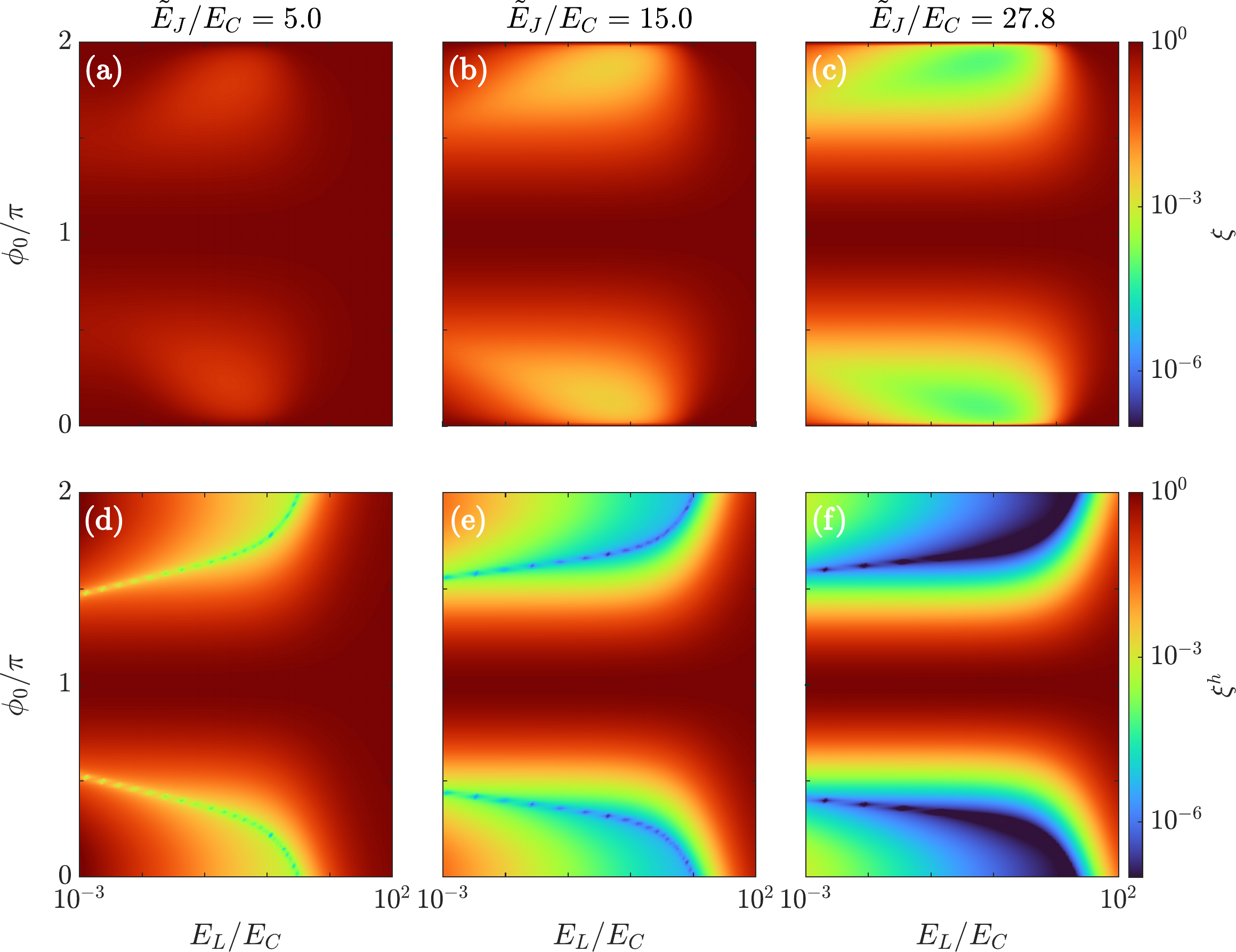}
    \caption{Comparison between exact overlaps $\xi$ from Eq. \eqref{eq:overlap_def}, top panels (a-c), and approximate solution $\xi^h$ from Eq. \eqref{eq_overlaph_approx}, bottom panels (d-f), against the ratio $E_L/E_C$ and the angle $\phi_0$ in Eq. \eqref{phase-minima}. 
 While the approximate solutions in (d-f) qualitatively capture the minimum overlap regions of the complete solutions (a-c), they exhibit significant quantitative discrepancies, with values differing by up to two orders of magnitude.}
    \label{fig_9}
\end{figure}

Since $E_S$ describes fluxon tunneling, the limit  of large detuning $|h_\sigma|\gg E_S$ describes very localized wave functions, with the sign of $h_{\sigma}$
fixing the spin. Specifically, the limit $h_{+}\rightarrow -\infty$ with $\theta_+=\theta\rightarrow\pi$ corresponds to a ground state  $\psi_{g+} \rightarrow \psi_{+}^0$. 
Equivalently, $h_{-}\rightarrow \infty$ with $\theta_-=\pi-\theta\rightarrow0$ corresponds to a ground state $\psi_{g-} \rightarrow \psi_{-}^1$. In this regime with negligible fluxon tunneling, the ground state overlap reduces to \eqref{eq:xi}, while
the first correction to this sweet spot near $\theta\approx \pi$ reads

\begin{gather}
\label{eq_overlaph_approx}
\xi^{h} \sim  \frac{ S_{0+0,1-0}+\frac{E_S}{\left|h_\sigma\right|} S_{0+0,0-0}  }{1-S\frac{E_S}{|h_{\sigma}|}}.
\end{gather}
\begin{figure}[t!]
    \centering
    \includegraphics[width=0.85\columnwidth]{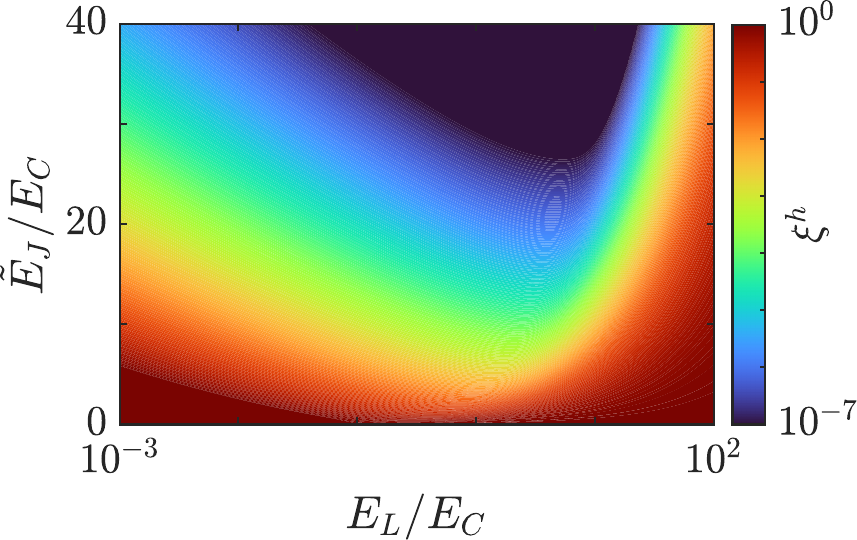}
    \caption{Approximate overlap $\xi^h$ from Eq.~\eqref{eq_overlaph_approx} versus $\tilde{E}_J/E_C$ and $E_L/E_C$. Overall, this harmonic approximation is in good qualitative agreement with the exact overlaps from Eq.~\eqref{eq:overlap_def} compare with panel
    Fig.~\ref{fig3}(e).}
    \label{fig_10}
\end{figure}
The phase slip semiclassical expression \eqref{eq:fluxon_tunnel} is only valid in the  $\tilde{E}_J\gg E_L$ limit. It is worth mentioning that the harmonic approximation disregards the potential non-gaussian contributions to tails of the wavefunctions, detail that can be critical in the regime $E_L/E_C \ll 1$. The non gaussian behavior of the wavefunction leads to a discrepancy between the calculated overlaps with the exact wavefunction and the ones determined using the harmonic approximation in Eq.~\eqref{eq_overlaph_approx}, see Fig.~\ref{fig_9} for a comparison. Nevertheless, the harmonic approximation qualitatively matches the behavior of the overlap, including the position of the minima. Furthermore, the approximation correctly describes the decrease on wavefunction overlap with $\tilde{E}_J/E_C$ and for $E_L/E_C\sim10$, see comparison between Fig.~\ref{fig_10} and Fig.~\ref{fig3}(e).
Similar good qualitative behavior over a large range of $\tilde{E}_J/E_C$ and $E_L/E_C$ can be found (Fig. \ref{fig_10}), which establishes the harmonic approximation as an excellent starting point for building physical intuition regarding the IPA qubit, as it remains remarkably accurate across a broad range of experimental parameters.

\bibliography{main.bib}
\end{document}